\documentclass[11pt]{article}
\pdfoutput=1
\usepackage{fullpage}
\usepackage{multirow}
\usepackage{cite}
\usepackage{graphicx}
\usepackage{amsmath}
\usepackage{amssymb}
\usepackage{cancel}
\title{}
\date{}
\newcommand{\ii}{\text{i}}

\begin{document}
\bibliographystyle{utphys}
\def\beq{\begin{equation}}
\def\eeq{\end{equation}}
\def\e{\epsilon}
\def\bsp#1\esp{\begin{split}#1\end{split}}
\def\beqa{\begin{eqnarray}}
\def\eeqa{\end{eqnarray}}
\def\eqn#1{eq.~(\ref{#1})}
\newcommand{\msbar}{\ensuremath{\overline{\text{MS}}}}
\newcommand{\DIS}{\ensuremath{\text{DIS}}}
\newcommand{\abar}{\ensuremath{\bar{\alpha}_S}}
\newcommand{\bb}{\ensuremath{\bar{\beta}_0}}
\newcommand{\rc}{\ensuremath{r_{\text{cut}}}}
\newcommand{\Nd}{\ensuremath{N_{\text{d.o.f.}}}}
\setlength{\parindent}{0pt}

\titlepage

\begin{flushright}
Edinburgh 2013/07 \\
\end{flushright}

\vspace*{0.5cm}

\begin{center}
{\large \bf The Non-Abelian Exponentiation theorem for multiple Wilson lines}

\vspace*{1cm}
\textsc{Einan Gardi$^a$\footnote{Einan.Gardi@ed.ac.uk}, Jennifer M. Smillie$^a$\footnote{J.M.Smillie@ed.ac.uk}
and Chris D. White$^b$\footnote{Christopher.White@glasgow.ac.uk} } \\

\vspace*{0.5cm} $^a$ Higgs Centre for Theoretical Physics, School of Physics and Astronomy, \\
The University of Edinburgh, Edinburgh EH9 3JZ, Scotland, UK \\ 

\vspace*{0.5cm} $^b$ SUPA, School of Physics and Astronomy, \\University of Glasgow, Glasgow G12 8QQ, Scotland, UK\\

\end{center}

\vspace*{0.5cm}

\begin{abstract}
We study the structure of soft gluon corrections to multi-leg scattering amplitudes in a non-Abelian gauge theory by analysing the corresponding product of semi-infinite Wilson lines. We prove that diagrams exponentiate such that the colour factors in the exponent are fully connected. 
This completes the generalisation of the non-Abelian exponentiation theorem, previously proven in the case of a Wilson loop, to the case of multiple Wilson lines in arbitrary representations of the colour group. 
Our proof is based on the replica trick in conjunction with a new formalism where multiple emissions from a Wilson line are described by effective vertices, each having a connected colour factor. The exponent consists of connected graphs made out of these vertices. We show that this readily provides a general colour basis for webs. We further discuss the kinematic combinations that accompany each connected colour factor, and explicitly catalogue all three-loop examples, as necessary for a direct computation of the soft anomalous dimension at this order. 
\end{abstract}

\vspace*{0.5cm}
\newpage
\tableofcontents

\section{Introduction}
Wilson lines in gauge theories have been studied for many years~\cite{Arefeva:1980zd,Polyakov:1980ca,Dotsenko:1979wb,Brandt:1981kf,Korchemsky:1985xj,Ivanov:1985np,Korchemsky:1985xu,Korchemsky:1985ts,Korchemsky:1986fj,Korchemsky:1987wg,Korchemsky:1988hd,Korchemsky:1988si,Collins:1989bt,Korchemsky:1991zp,Kidonakis:1998nf,Kidonakis:1997gm,Kidonakis:1996aq}, and continue to be investigated in both gauge theories \cite{Drummond:2007cf,Basso:2007wd,Alday:2007hr,Pestun:2007rz,Drukker:2012de,Chien:2011wz,Cherednikov:2012qq,Cherednikov:2012yd,Henn:2013wfa} and gravity~\cite{Naculich:2011ry,White:2011yy,Akhoury:2011kq,Miller:2012an}. In QCD, they are of interest for collider phenomenology, as they govern the structure of soft gluon contributions in hard scattering processes to all orders in perturbation theory. Important applications include Sudakov and transverse-momentum resummation \cite{Collins:1989gx,Korchemsky:1992xv,Korchemsky:1993uz,Catani:1996yz,Kidonakis:1998nf,Kidonakis:1997gm,Kidonakis:1996aq,Oderda:1999kr,Bonciani:1998vc,Beneke:2009rj,Beneke:2009ye,Ahrens:2010zv,Czakon:2013goa,Bauer:2000ew,Bauer:2000yr,Bauer:2001ct,Bauer:2001yt,Bauer:2002nz,Becher:2006nr,Becher:2006mr,Becher:2007ty,Yennie:1961ad,Sterman:1986aj,Catani:1989ne,Laenen:2008gt} as well as analysis of the high-energy 
limit~\cite{Sotiropoulos:1993rd,Korchemsky:1993hr,Korchemskaya:1994qp,Korchemskaya:1996je,Balitsky:1995ub,Kovchegov:1996ty,Balitsky:2001gj,Balitsky:2009yp,JalilianMarian:1996xn,Gardi:2006rp,DelDuca:2011xm,DelDuca:2011ae,Kovchegov:2012mbw,Mueller:1993rr}.\\

Wilson lines, or path-ordered exponentials, universally emerge upon taking the soft approximation. To analyse the singularities of a given QCD amplitude involving $L$ hard partons ${\cal M} (p_1,p_2,\ldots, p_L)$ one may consider the corresponding correlator of semi-infinite Wilson line rays:  
\begin{equation}
\label{soft_correlator}
Z=\Big<\phi_{\beta_1}\,\otimes\,\phi_{\beta_2}\,\otimes\,\ldots\,\,\otimes\,\phi_{\beta_L}\Big>
\end{equation}
where each $\phi_{\beta_i}$ is a Wilson line stretching from the interaction point to infinity along the classical trajectory of parton $i$ ($\beta_i$ is a four-velocity vector in the direction of $p_i$):
\begin{equation}
\label{phi_def}
\phi_{\beta_i}={\cal P} \exp\left[\ii g_s\int_0^{\infty}ds\beta_i^\mu\,A_\mu(s\beta_i)\right].
\end{equation}
Here $A_{\mu}=A_{\mu}^aT_i^a$ is a gauge field in the representation of the parton $i$, and ${\cal P}$ denotes path-ordering of the colour matrices along the line. In a factorization of an amplitude or a cross section, the correlator $Z$ captures all soft gluon interactions with the hard scattered partons. 
It is the incoherence between the soft interactions and the hard interaction on the one hand and the collinear ones (incoming partons or final-state jets) on the other, which ultimately leads to factorization of the soft subprocess, making the process-independent calculation of soft interactions and the associated singularities (or logarithms) possible.\\

The further advantage of using the Wilson line correlators $Z$ stems from the
fact that their \emph{ultraviolet} renormalisation contains the structure of infrared singularities in the corresponding scattering amplitudes~\cite{Ivanov:1985np,Korchemsky:1985xj,Korchemsky:1985xu,Korchemsky:1985ts,Korchemsky:1986fj,Korchemsky:1987wg,Korchemsky:1991zp,Kidonakis:1998nf,Kidonakis:1997gm,Kidonakis:1996aq}. The singularities are encoded in the \emph{soft anomalous dimension}, defined through the scale dependence of $Z$. The soft anomalous dimension captures the full complexity of the coherent process of non-collinear soft interactions depending on the underlying hard kinematics and colour flow.
Of particular interest are scattering amplitudes involving four or more hard coloured partons, where the soft anomalous dimension is a matrix in colour-flow space.  These matrices have been computed at leading and next-to-leading orders and their physical implications have been studied in detail, starting with the early works of refs.~\cite{Korchemsky:1993hr,Sotiropoulos:1993rd,Korchemskaya:1994qp,Korchemskaya:1996je,Oderda:1999kr,Kidonakis:1998nf,Kidonakis:1997gm,Kidonakis:1996aq} and more recently in refs.~\cite{Dokshitzer:2005ek,Seymour:2005ze,Kyrieleis:2005dt,Forshaw:2006fk,Kidonakis:2009ev,Kidonakis:2010dk,Mitov:2009sv,Becher:2009kw,Becher:2009cu,Gardi:2009qi,Becher:2009qa,Gardi:2009zv,Aybat:2006wq,Aybat:2006mz,Dixon:2009ur,DelDuca:2011ae,DelDuca:2011xm,Vernazza:2011aa,Sjodahl:2009wx,Catani:2011st,Catani:2012iw,Forshaw:2012bi,Beneke:2009rj,Czakon:2009zw,Beneke:2009ye,Ferroglia:2009ep,Ferroglia:2009ii,Chiu:2009mg,Mitov:2010xw,Ferroglia:2010mi,Bierenbaum:2011gg}. \\

Wilson line correlators are significantly simpler than the amplitude, as they do not involve spin degrees of freedom nor do they depend on the energy scale of the scattered partons. This leads to symmetries, notably eq.~(\ref{phi_def}) is invariant with respect to velocity rescaling.
A detailed analysis of these properties has led to a recent conjecture for the complete all-order perturbative structure of infrared singularities in \emph{massless} QCD amplitudes, where the Wilson lines are
 lightlike ($\beta_i^2=0$). This result is known as the \emph{dipole formula}~\cite{Becher:2009cu,Gardi:2009qi,Becher:2009qa,Gardi:2009zv}, as it implies that all infrared singularities are generated by an exponent containing colour correlations between at most pairs of partons (dipoles). The dipole formula is known to yield the exact result up to two loop order, both from general considerations and from explicit calculations~\cite{Aybat:2006wq,Aybat:2006mz}. 
Possible corrections to the soft anomalous dimension going beyond the dipole formula begin at three loops, and have been investigated in refs.~\cite{Becher:2009kw,Dixon:2009ur,DelDuca:2011ae,DelDuca:2011xm,Vernazza:2011aa,Naculich:2013xa}, where stringent constraints have been found, but a three-loop calculation has not yet been done.\\

Wilson lines are also useful to study long-distance singularities of amplitudes involving heavy partons, such as top quarks. In this case the velocities are timelike, $\beta_i^2>0$. Also here the soft anomalous dimension has been computed through two loops~\cite{Kidonakis:2009ev,Kidonakis:2010dk,Mitov:2009sv,Becher:2009kw,Beneke:2009rj,Czakon:2009zw,Beneke:2009ye,Ferroglia:2009ep,Ferroglia:2009ii,Chiu:2009mg,Mitov:2010xw,Ferroglia:2010mi,Bierenbaum:2011gg}. In contrast to the massless-parton case, here three-parton correlations involving colour and kinematic degrees of freedom do appear, and it is expected that similar, multi-parton correlations involving more partons would appear at higher orders. \\

The long-distance singularity structure of amplitudes for both massless and massive partons is significantly simpler than finite corrections to these amplitudes, and reveals an interesting interplay between colour and kinematics, which  may be related to other such relationships discussed recently~\cite{Bern:2008qj} (see~\cite{Oxburgh:2012zr} for a discussion relating to the soft limit). Furthermore, the general structure of infrared singularities is similar for a variety of non-Abelian gauge theories and has already been used (e.g. in~\cite{Bern:2005iz}) in understanding the all order structure of ${\cal N}=4$ amplitudes at the planar level.
This provides a strong theoretical motivation, going well beyond the immediate application to collider jet physics, to study these singularities at higher loop orders, and for general~$N_c$~\cite{Naculich:2011my,Naculich:2013xa,Naculich:2011pd,Naculich:2008ys}, with the long term goal of determining the anomalous dimensions to all loops and for strong coupling. \\

In hard interaction processes involving two coloured lines, soft gluon interactions are captured by a vacuum expectation value of a product of two semi-infinite Wilson line operators with opposite colour charges, 
$Z=\left<\phi_{\beta_1}\,\phi_{\beta_2}\right>$, or equivalently a cusped Wilson loop closing at infinity. In this case it has been known for many years that the \emph{exponent} $w$, defined by $Z=\exp[w]$, can itself be given a Feynman diagram interpretation and thus be computed directly\footnote{A pedagogical review of this material can be found in \cite{Berger:2003zh}. }~\cite{Gatheral:1983cz,Frenkel:1984pz,Sterman:1981jc} where only certain diagrams contribute: so-called \emph{webs} \cite{Sterman:1981jc}. 
In QED, webs are connected subdiagrams\footnote{By subdiagram we refer to the (soft) gauge field components of the diagram, excluding the Wilson lines themselves. }, a result first shown in~\cite{Yennie:1961ad}. 
In QCD, for a Wilson loop, webs are \emph{irreducible} subdiagrams,
namely those diagrams whose colour factors cannot be written as a product of the colour factors of separate subdiagrams. An equivalent topological criterion is that constituent parts of an irreducible diagram cannot be disconnected by merely cutting through the two Wilson lines~\cite{Gatheral:1983cz,Frenkel:1984pz,Sterman:1981jc}. This simple criterion is sufficient to classify all relevant diagrams in the exponent $w$ to all orders. 
Furthermore, the colour factor $\widetilde{C}(D)$ which accompanies any such diagram $D$ in the exponent -- referred to below as the Exponentiated Colour Factor (ECF) -- is not the same as the conventional colour factor $C(D)$ of that diagram: it is the \emph{maximally non-Abelian} part of this colour factor.
This means that the ECF of any web corresponds to diagrams in which emitted gluons are \emph{fully connected} by multiple gluon vertices away from the Wilson lines. This is known as the \emph{non-Abelian exponentiation theorem}, and has been applied in Wilson line calculations in various contexts. \\

The notion of webs has recently been generalised to multi-parton scattering or, equivalently, to vacuum expectation values of any number of Wilson lines meeting at a point~\cite{Mitov:2010rp,Gardi:2010rn}, eq.~(\ref{soft_correlator}) above. 
Similarly to the two-parton case, the exponent $w$, defined through $Z=\exp[w]$, can be given a direct diagrammatic interpretation. 
Due to the non-trivial colour exchange at the hard interaction vertex, infrared singularities in multi-parton scattering become highly non-trivial, and exhibit a rich mathematical structure. 
A major difference to the two parton (Wilson loop) case is that
 webs, instead of being single irreducible diagrams, are closed sets of (possibly reducible) diagrams related by gluon permutations on the Wilson lines. Each set of diagrams constitutes a single web $W$, whose contribution to the exponent $w$ has the form 
\begin{equation}
W=\sum_{D}{\cal F}(D)\widetilde{C}(D)=\sum_{D,D'}{\cal F}(D)R_{DD'}C(D').
\label{Wform}
\end{equation}
Here $\{{\cal F}(D)\}$ and $\{C(D)\}$ are, respectively, the sets of kinematic and colour parts associated with all diagrams $D\in W$, diagrams which are related to each other by permuting the order of gluon attachments to the Wilson lines, while $R_{DD'}$ is a matrix of rational numbers called the \emph{web mixing matrix}. Each web has an associated web mixing matrix, and they encode a large amount of physics: web mixing matrices dictate how colour and kinematic information is entangled in the exponent of the soft gluon amplitude. These matrices were further studied in~\cite{Gardi:2011wa,Gardi:2011yz}, which established some general properties. One basic property that will be important below, is that web mixing matrices are idempotent, namely they act as projection operators, selecting particular linear combinations of colour factors $\{C(D)\}$ to appear in the exponent.  We will return to characterise these combinations below.
The mixing matrix can also be viewed as acting on the kinematic factors, generating particular linear combinations of  $\{{\cal F}(D)\}$ in which certain subdivergences cancel, as dictated by the renormalisation properties of the vertex at which the Wilson lines meet \cite{Gardi:2011yz}. 
It is also known that the contents of web mixing matrices can be obtained purely from combinatoric reasoning~\cite{Gardi:2011wa}, which has been further investigated in a mathematical context~\cite{Dukes:2013wa,Dukes:2013gea}. \\

Despite the above progress, much remains unknown about both the physics and combinatorics of webs. In addition, a number of conjectures and ideas regarding multiparton webs (and their similarity to the canonical two-parton webs of refs.~\cite{Gatheral:1983cz,Frenkel:1984pz,Sterman:1981jc}) are unproven. Chief among these is the question of whether the maximal non-Abelian (or ``fully connected'') nature of colour factors in the exponent of the soft amplitude generalises from the two-parton to the multi-parton case. One might expect that this property would generalise, and this has been assumed in recent studies of infrared singularities (e.g.~\cite{Becher:2009qa,Dixon:2009ur,Ahrens:2012qz}, which considers possible corrections to the QCD dipole formula). 
However, given that the topological criterion for identifying webs as irreducible diagrams does not carry over from the Wilson loop case to the general case of several Wilson lines -- indeed in the latter case reducible diagrams do contribute to the exponent -- it is not at all obvious if the ``fully connected'' nature of colour factors would be realised.  This calls for a general proof of this result, which is the subject of this paper. Furthermore, progress in computing infrared singularities requires determining the relevant kinematic combinations accompanying each connected colour factor. We will proceed to identify those in this paper. \\

The structure of the paper is as follows. In the next section we prove that all colour factors in the exponent of a correlation function involving any number of Wilson-lines are fully connected. Our proof relies on a variant of the replica trick argument adopted
in refs.~\cite{Laenen:2008gt,Gardi:2010rn,Gardi:2011wa}. Before setting up the proof we use a simple two-loop example to illustrate the idea in section~\ref{sec:ex}. 
In order to formulate the proof,  in section~\ref{sec:proof} we introduce a new formalism where the interaction of multiple gluons with a Wilson line is represented by effective vertices.
We then show that the exponent is obtained by summing up connected graphs made of these vertices. The explicit form of the kinematic and replica-number dependence of these vertices in the first three orders is presented in section~\ref{sec:vertices}. In section~\ref{sec:basis} we further show that the vertex-based formalism provides a natural colour basis for webs.
The use of this method in calculations is explained and illustrated in
section~\ref{sec:effect-vert-pract}.  Finally, in section~\ref{sec:3loop}, we classify
all possible examples of webs at three-loop order which connect three or four Wilson lines, determine the connected colour factors in a common basis, and derive the combinations of kinematic integrals that accompany each connected colour factor. This may be used in a
calculation of the multiparton soft anomalous dimension at this order. In section~\ref{sec:conclude} we briefly discuss our results and conclude. In a forthcoming publication we undertake a combinatoric investigation which provides further insight into webs and their connected colour factors~\cite{Dukes:2013gea}.

\section{The connected colour factor theorem}
\label{sec:theorem}

In this section we prove the following result:
\newtheorem{theorem}{Theorem}[section]
\begin{theorem}
\label{theorem}
Radiative corrections to correlators of any number of Wilson lines in arbitrary representations of the gauge group exponentiate such that the colour factors appearing in the exponent all correspond to connected graphs. \\
\end{theorem}
This generalises the non-Abelian exponentiation theorem of refs.~\cite{Gatheral:1983cz,Frenkel:1984pz} from the Wilson loop case to an arbitrary product of Wilson lines. 
Before presenting the formalism and the proof we provide a simple example to illustrate the idea.

\subsection{Connected colour factors -- a simple example}
\label{sec:ex}
\begin{figure}[htb]
\begin{center}
\vspace*{10pt}
\begin{minipage}[b]{0.55\linewidth}
\scalebox{.7}{\includegraphics{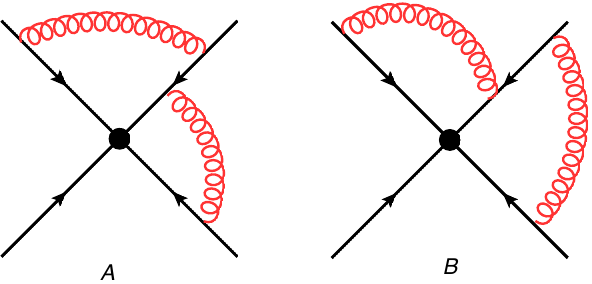}}
\end{minipage}
\begin{minipage}[b]{0.15\linewidth}
\begin{tabular}{c}
\scalebox{.4}{\includegraphics{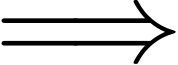}}
\\
\\
\\
\\
\\
\\
\,
\end{tabular}
\end{minipage}
\begin{minipage}[b]{0.1\linewidth}
\scalebox{0.72}{\includegraphics{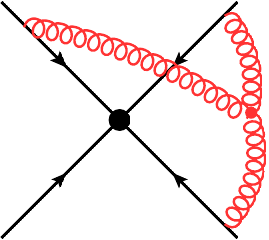}}
\end{minipage}
\vspace*{-40pt}
\caption{2-loop web connecting 3 Wilson lines and its exponentiated colour factor.}
\label{fig:121}
\end{center}
\end{figure}

Let us begin by considering a simple example of a multiparton web in order to introduce the notion of a connected colour factor. We consider the web shown on the left-hand side of figure~\ref{fig:121}, and consisting of two separate gluon emissions between the Wilson lines. This is the simplest non-trivial example of a multiparton web. 
In characterising webs, we
adopt the notation\footnote{Note that this notation does not fully specify the web due to the possibility of interactions off the Wilson lines. A more refined notation will be introduced where necessary.} introduced in~\cite{Gardi:2010rn}, where $(n_1,n_2,\ldots
n_L)$ denotes a web with $n_i$ gluon emissions on Wilson line $i$. Using this notation the web of figure~\ref{fig:121} will be denoted $W_{(1,2,1)}$.
Following eq.~(\ref{Wform}), the contribution of this web to the exponent of the scattering amplitude is 
\[
W_{(1,2,1)}= {\cal F}(D) R^{{(1,2,1)}}_{D,D^{\prime}} C({D^{\prime}}),
\]
where $D, D^{\prime}$ run over the diagrams in the set, and $R^{{(1,2,1)}}_{D,D^{\prime}}$ is the appropriate web-mixing matrix. In this particular case, this takes the form~\cite{Gardi:2010rn}:
\begin{align}
\begin{split}
W_{(1,2,1)}&=
\left[ \begin {array}{c} {\cal F}(A)\\ \noalign{\medskip}{\cal F}(B)\end {array} \right]^{T}  \left[ \begin {array}{rr} \frac12&-\frac12\\ \noalign{\medskip}-\frac12&\frac12\end {array}
 \right] 
 \left[ \begin {array}{c} C(A)\\ \noalign{\medskip}C(B)\end {array} \right] 
\\
&=  \frac12 \Big({\cal F}(A)-{\cal F}(B)\Big)\Big(C(A)-C(B)\Big)
\\&= \Big({\cal F}(A)-{\cal F}(B)\Big)\widetilde{C}(A).
\end{split}
\label{R121}
\end{align}
We see that in this case only a single combination of kinematic factors, the difference between the two integrals ${\cal F}(A)-{\cal F}(B)$, contributes to the exponent of the soft amplitude. Furthermore, it does so with the colour factor
\begin{equation}
\widetilde{C}(A)=\frac{1}{2}\Big(C(A)-C(B)\Big),
\label{ECFA}
\end{equation}
which is the exponentiated colour factor (ECF) associated with diagram $A$, adopting the notation of~\cite{Gardi:2010rn}. We may go further than this by writing down the colour factors explicitly:
\begin{equation}
\label{C_AB}
C(A)=T_1^{a}T_2^{ab} T_3^{b}\,,\qquad C(B)=T_1^{a}T_2^{ba} T_3^{b}
\end{equation}
 where we used the shorthand notation:
\begin{equation}
T_i^{abcd...}=T_i^aT_i^bT_i^cT_i^d\ldots,
\end{equation}
and $T_i^a$ is a generator in the representation associated with Wilson line $i$~\cite{Catani:1996vz} corresponding to the emission of a gluon of adjoint index $a$. Note that we have ordered the colour matrices (on line 2) inwards towards the hard interaction, consistent with the path-ordering implied by the directions of the Wilson lines in figure~\ref{fig:121}. 
We emphasize that in eq.~(\ref{C_AB}) ordering is relevant only among colour matrices corresponding to a given line; as indicated in eq.~(\ref{soft_correlator}) we always consider an external product between the different lines, which may each belong to a different representation.\\

Using the colour algebra on Wilson line 2
\begin{equation}
\label{algebra}
\left[T_2^{a},T_2^{b}\right] ={\rm i}f^{abc} 
T_2^{c},
\end{equation}
we may rewrite the ECF obtained above as
\begin{equation}
  \label{eq:finalctilde}
\widetilde{C}(A)=\frac12 T_1^{a}\left[T_2^{a},T_2^{b}\right] T_3^{b}=\frac12 {\rm i}f^{abc} 
T_1^{a}T_2^{c}T_3^{b}.
\end{equation}
We thus see that the only colour factor entering the exponent in this case, corresponds (up to an overall normalisation constant) with the non-Abelian diagram on the right-hand side of figure~\ref{fig:121}. In contrast with diagrams $A$ and $B$, which upon removing the Wilson lines remain non-connected, the ECF corresponds to a fully connected diagram: the number of gluon emissions on the Wilson lines is reduced, such that the gluons are joined by three-gluon vertices off the Wilson lines. This illustrates theorem \ref{theorem}. In the next subsection we will prove that it holds in general, for any number of Wilson lines and to any order in perturbation theory, thus generalizing the non-Abelian exponentiation theorem~\cite{Gatheral:1983cz,Frenkel:1984pz} to the multiparton case.\\

A brief historical comment is due here, which will also clarify the terminology we use. ECFs were first characterized as  ``\emph{maximally non-Abelian}'' by Gatheral~\cite{Gatheral:1983cz}. This is precise when considering a single Wilson loop, and so long as only quadratic Casimirs are relevant\footnote{In this case the colour factor in the exponent for a Wilson loop in representation $R$, takes the form $C_R C_A^{K-1}\, \alpha_s^K$, where $C_R$ is the quadratic Casimir in representation $R$ and $C_A$ corresponds to the Adjoint representation.}; this holds through three loops, but becomes ambiguous beyond this order. Frenkel and Taylor \cite{Frenkel:1984pz} characterized the nature of ECFs for a Wilson loop more generally as the ``\emph{connected part}'' of the ordinary colour factor, and provided an algorithm to determine it by unwinding all crossings among exchanged gluons using the colour algebra. 
This algorithm, however, does not apply to the case of multiple Wilson lines
where, for example, reducible diagrams (where the gluons may not cross each
other at all) do contribute to the exponent~\cite{Mitov:2010rp,Gardi:2010rn}. We
nevertheless observe here that the special nature of ECFs is preserved: they all
correspond to \emph{fully connected graphs}. A general algorithm to determine
the ECFs\footnote{See also  Ref.~\cite{Mitov:2010rp} where an alternative
  approach to diagrammatic exponentiation was presented.} using web mixing
matrices was established in refs.~\cite{Gardi:2010rn,Gardi:2011wa}.  The formalism we develop below provides an alternative method to compute them, which in particular allows us to prove that they are all fully connected.
\\

Some additional comments are in order regarding the two-loop example above, and its generalisation to higher orders. We saw that there was only one connected colour factor in this case. This can be understood as follows. Web mixing matrices are idempotent\footnote{This property was first observed in~\cite{Gardi:2010rn}, and then proven in~\cite{Gardi:2011wa}.} and thus have eigenvalues 0 and~1, with some potential multiplicity. One may introduce a diagonalizing matrix $Y$ whose rows are the left-eigenvectors of $R$, and rewrite eq.~(\ref{Wform}) as
\begin{align}
W&={\cal F}^{\rm T}\,R\,C\notag\\
&=\left({\cal F}^{\rm T}\,Y^{-1}\right)Y\,R\,Y^{-1}\left(YC\right)\notag\\
&=\left({\cal F}^{\rm T}\,Y^{-1}\right)\,{\rm diag}(\lambda_1,\lambda_2,\ldots
,\lambda_d)\left(YC\right),
\label{Wform2}
\end{align}
where $\{\lambda_i\}$ are the eigenvalues of $R$ and $d$ is its dimension. Without loss of generality, we choose the first $r$ eigenvectors in $Y$ (where $r$ is the rank of $R$) to be the ones corresponding to eigenvalue~1, and the other $d-r$ eigenvectors to correspond to the zero eigenvalue. We further know, based on the zero sum-row property of $R$ proven in~\cite{Gardi:2011wa}, that there is at least one eigenvalue zero, so $d-r\geq 1$. 
One then obtains
\begin{equation}
W=\sum_{H=1}^r\left({\cal F}^{\rm T}\,Y^{-1}\right)_H\,\left(YC\right)_H,
\label{Wform3}
\end{equation}
with $r<d$, where all contributions associated with zero eigenvalues of $R$ have been manifestly projected out. The intepretation of eq.~(\ref{Wform3}) is as follows. Each left eigenvector of $R$ of unit eigenvalue is associated with an effective colour factor, found by contracting the relevant row of $Y$ (indexed $H$) with the vector of colour factors $C$. A given web $W$ has $r$ such colour factors which are mutually independent.
Each effective colour factor is associated with a given combination of kinematic factors, found by contracting the vector of kinematic factors ${\cal F}$ with the corresponding column of $Y^{-1}$. \\

In the two-loop example considered above, there is only one left eigenvector of the web mixing matrix with unit eigenvalue. This can be seen from the fact that the two rows of the mixing matrix in eq.~(\ref{R121}) are not independent of each other (which follows from the general property that all mixing matrices have at least one eigenvalue zero). 
There is thus a single effective colour factor, associated with a single combination of kinematic factors. We have seen that this colour factor is fully connected, and we will see that the generalisation of this to higher orders is the following: left-eigenvectors of web mixing matrices, of unit eigenvalue, give exponentiated colour factors which are fully connected. The combination of kinematic factors which accompanies each colour factor can be found by explicitly constructing the matrix $Y$ of left-eigenvectors of $R$, and applying this as in eq.~(\ref{Wform3}). \\

We emphasize that, for a given web, the basis of $r$ fully connected colour factors is not unique. The reason for this is that colour factors may be interrelated by Jacobi identities, and we will see examples of this in section~\ref{sec:3loop}. To proceed one should therefore construct a suitable basis of independent connected colour factors at any given order. One of the advantages of the new effective vertex formalism we develop in this paper is that it provides a natural basis for these connected colour factors.\\

Having introduced the concept of connected colour factors, and related these to web mixing matrices, our next aim is to prove in general that colour factors which occur in the exponent are indeed all fully connected. This is the subject of the next subsection.

\subsection{Proof of the theorem}
\label{sec:proof}

In the previous section, we showed that the two-loop web of figure~\ref{fig:121}
gave rise to an exponentiated colour factor that is fully connected. We then saw
quite generally that a given web contributes a number $r$ of ECFs, where $r$ is
the rank of the corresponding web mixing matrix. We stated in theorem
\ref{theorem} that each such ECF is also a fully connected colour factor, and
the aim of this section is to prove this result. We will employ the replica trick of statistical physics, which has also been used in the present context in Refs.~\cite{Laenen:2008gt,Gardi:2010rn,Gardi:2011wa}. Let us first introduce the necessary formalism and recall the basic reasoning.\\

The Wilson line correlator of eq.~(\ref{soft_correlator}) can be written as a functional integral over the gauge field as follows: 
\begin{equation}
Z[J]=\int{\cal D}A_\mu e^{\ii S[A_\mu]}\prod_{l=1}^L {\cal P}\exp\left[\ii g_s\int dx_l^\mu\,A_\mu(x_l)\right],
\label{Zdef}
\end{equation}
where $S$ stands for the usual gauge theory action including sources:
\begin{equation}
S[A_\mu]=\int d^dx \Big( {\cal L}_{\rm YM} + J^{a\,\mu}(x) A^{a}_{\mu}(x)\Big)\,. 
\end{equation}
The line integral in each exponential factor is defined as in eq.~(\ref{phi_def}), and ${\cal P}$ denotes path-ordering of the colour matrices along the line. In eq.~(\ref{soft_correlator}) these are straight, semi-infinite lines extending along the classical trajectories of the scattered partons in the corresponding amplitude; similarly, their colour representations are inherited from the respective partons. These details, however, play no role in establishing the connected nature of the ECF below, and thus we shall keep the notation general. Note that $A_\mu$ is a matrix-valued non-Abelian Yang-Mills field. Each path-ordered exponential $l$ corresponds to a given representation where $A_\mu=A_\mu^a T_l^a$ but we will not need to specify these, and keep the discussion general throughout the paper. \\  

As usual, when expanded perturbatively in the coupling the functional integral,
eq.~(\ref{Zdef}), generates all possible radiative corrections, which can be
described in terms of Feynman diagrams. Rather than considering this expansion,
we instead directly construct the diagrammatic description of the exponent $w$ defined via
\begin{equation}
Z=\exp\left[w\right]\,.
\label{Z_w}
\end{equation}
To this end, consider first a theory which has $N$ identical, non-interacting copies (replicas) of the gauge field $A_\mu$. 
The generating function for such a replicated theory is given by
\begin{align}
\label{Zrepdef}
Z_{\rm rep.}\left[\left\{J^{a \mu (i)}\right\}\right]&\equiv\int{\cal D}A^{(1)}_\mu\,{\cal D}A^{(2)}_\mu\ldots{\cal D}A^{(N)}_\mu\,\, e^{\ii\sum_{i=1}^NS[A^{(i)}_\mu]}\notag\\
&\quad\times\prod_{l=1}^L {\cal P}\exp\left[\ii g_s\int dx_l^\mu\,A^{(1)}_\mu(x_l)\right]\ldots{\cal P}\exp\left[\ii g_s\int dx_l^\mu\,A^{(N)}_\mu(x_l)\right]\,,
\end{align}
where we have also replicated the Wilson lines, such that each line $l$ sources all replicas. 
The corresponding path-ordered exponentials are ordered such that the leftmost one corresponds to the field of replica $A^{(1)}$, the next to $A^{(2)}$, and so on, ending with the last replica $A^{(N)}$ on the right.
It should be emphasised that the matter content of the theory is also replicated keeping the replicas independent, and so are the sources $J^{a \mu (i)}$ for each gauge field replica $(i)$. 
Upon identifying the sources $J^{a \mu (i)}=J^{a \mu}$ for all replicas one obtains an obvious yet very useful relation between the replicated theory and the original one
\begin{align}
Z_{\rm rep.}[J]=\left(Z[J]\right)^N,
\label{Zrepdef_rel}
\end{align}
where we have explicitly made use of the fact that the replicas are non-interacting.
By a simple mathematical identity, one has
\begin{equation}
Z^N=1+N\log Z+{\cal O}(N^2),
\label{ZN}
\end{equation}
from which it follows that the original generating functional satisfies
\begin{equation}
Z=\exp[w]=\exp\left[\sum_W W\right],
\label{Zexp}
\end{equation}
where the sum is over all diagrams $W$ which are ${\cal O}(N^1)$ in the replicated theory.\\

Reference~\cite{Gardi:2010rn} proceeded by noting that the product of Wilson line operators on each external line in eq.~(\ref{Zrepdef}) is such that the replica number is increasing on each line, so that the generating functional for the replicated theory may be rewritten as
\begin{equation}
\!Z_{\rm rep.}\left[\left\{J^{a \mu (i)}\right\}\right]=\int\!{\cal D}A^{(1)}_\mu{\cal D}A^{(2)}_\mu\ldots{\cal D}A^{(N)}_\mu e^{\ii\sum_{i=1}^NS[A^{(i)}_\mu]}
\prod_{l=1}^L {\cal R}{\cal P}\exp\left[\ii g_s\sum_{j=1}^N\int dx_l^\mu\,A^{(j)}_\mu(x_l)\right].
\label{Zrepdef2}
\end{equation}
Here ${\cal R}$ is a replica-ordering operator, which overrides the path-ordering for gluon emissions whose replica numbers are different, so that the latter are increasing along the Wilson line, preserving the order in eq.~(\ref{Zrepdef}). 
Using eqs.~(\ref{Zrepdef_rel}) and (\ref{ZN}), the Feynman diagram structure of the exponent $w$ can then be obtained upon taking the ${\cal O}(N^1)$ coefficient of the diagrams generated by eq.~(\ref{Zrepdef2}).
In this calculation, the presence of the ${\cal R}$ operator means that the colour factors in the replicated theory differ from those in the original theory, and it is this that ultimately leads to the structure of ECFs in terms of web-mixing matrices of refs.~\cite{Gardi:2010rn,Gardi:2011wa,Gardi:2011yz}.
\\

It is convenient for our present purposes to implement the ${\cal R}$ and ${\cal
  P}$ operators within the exponent itself. To this end, we note that it is possible to represent a single path-ordered exponential as a conventional exponential, whose exponent contains an infinite series of terms~\cite{Grensing:1986bg}
\begin{align}
\label{Path_ordered_repeated_BCH}
{\cal P}\exp\left(\ii g_s \int_{s_0}^{s_f} ds A(s) \right)=\exp\left\{\sum_{K=1}^{\infty} (\ii g_s)^K F_K\right\}.
\end{align}
Here we have parametrised the path along the Wilson line using a parameter $s$, such that 
$x^{\mu}=x^{\mu}(s)$, and defined
\begin{equation}
A(s)\equiv \dot{x}^{\mu}A_{\mu}(x)= \dot{x}^{\mu}A_{\mu}^{a}(x)T^{a},
\label{Asparam}  
\end{equation}
where $\dot{x}^{\mu}\equiv dx^{\mu}/ds$.
The expansion coefficients $\{F_K\}$ that appear in eq.~(\ref{Path_ordered_repeated_BCH}), up to third order, are given by 
\begin{subequations}
\begin{align}
F_1&=\int_{s_0}^{s_f} ds A(s)\\
F_2&=\frac12 \int_{s_0}^{s_f} ds_1 \int_{s_0}^{s_f} ds_2\,\Theta(s_2-s_1)\, \left[A(s_2),A(s_1)\right]\,\\
\begin{split}
F_3&=\frac{1}{6} \int_{s_0}^{s_f} ds_1 \int_{s_0}^{s_f} ds_2\,\int_{s_0}^{s_f} ds_3\,\Theta(s_3-s_2)\,\Theta(s_2-s_1)\times\\&\qquad\qquad\times  
\Big(\left[\left[A(s_3),A(s_2)\right],A(s_1)\right]
- \left[\left[A(s_2),A(s_1)\right],A(s_3)\right]
\Big),
\end{split}
\end{align}
\end{subequations}
where $\Theta(s)$ is the Heaviside function and the square brackets represent commutators,
\[
\left[A(s_2),A(s_1)\right]=A(s_2)A(s_1)-A(s_1)A(s_2).
\]
Higher-order coefficients are progressively more complicated, but always have
the general form of \emph{a fully nested commutator} involving $A(s_i)$
functions evaluated at different points $s_i$, together with integrals over the
dummy variables $\{s_i\}$ with accompanying Heaviside functions\footnote{The
  derivation of eq.~(\ref{Path_ordered_repeated_BCH}) proceeds by discretising
  the Wilson line operator into a product of exponentials, and then combining
  these using the Baker-Campbell-Hausdorff (BCH) formula, before taking the
  continuum limit. It is the use of the BCH formula that results in fully nested
  commutator structures at each order - see ref.~\cite{Grensing:1986bg}.}. \\

Each Wilson line in the replicated theory, eq.~(\ref{Zrepdef}), carries a product of path-ordered exponentials - one for each replica, and each having the form of eq.~(\ref{Path_ordered_repeated_BCH}) if we are to express the path-ordered exponential in terms of a conventional exponential. The advantage of doing so is that one may then combine the Wilson line exponentials corresponding to the different replicas using multiple applications of the Baker-Campbell-Hausdorff (BCH) formula:
\begin{align}
\label{repeated_BCH}
{\rm e}^{x_1}{\rm e}^{x_2}\ldots {\rm e}^{x_N}=\exp\left\{\sum_{n=1}^{\infty} E_n\right\},
\end{align}
where the $\{E_n\}$, up to third order, are given by
\begin{subequations}
\begin{align}
E_1&=\sum_{i=1}^N x_i\\
E_2&=\frac12 \sum_{i<j} \left[x_i,x_j\right]\\
E_3&=\frac{1}{12} \sum_{i\neq j} \left[\left[x_i,x_j\right],x_j\right]
+\frac{1}{6}\sum_{i<j<k}  \left[\left[x_i,x_j\right],x_k\right]
-  \left[\left[x_j,x_k\right],x_i\right]
\,,
\end{align}
\end{subequations}
where $i$, $j$ and $k$ are replica indices.
Again, each higher-order term involves a series of fully nested commutators. To combine the $N$ Wilson-line operators on each Wilson line, one uses eq.~(\ref{repeated_BCH}) with 
\begin{displaymath}
x_i=\sum_{K=1}^{\infty} (\ii g_s)^K F^{(i)}_K,
\end{displaymath}
where the superscript of $F^{(i)}_K$ on the right-hand side indicates that the expansion coefficient of the path-ordered exponent involves the gauge field with replica index $i$. Putting things together, one finds that the product of Wilson-line operators associated with a given external line has the form
 \begin{align}
\label{Replicated_Path_ordered_repeated_BCH}
\begin{split}
&
{\cal P}\exp\left(\ii g_s \int_{s_0}^{s_f} \!\! ds A_1(s) \right)
{\cal P}\exp\left(\ii g_s \int_{s_0}^{s_f} \!\! ds A_2(s) \right)
\,\ldots\,
{\cal P}\exp\left(\ii g_s \int_{s_0}^{s_f} \!\! ds A_N(s) \right)
\\
\equiv &
\exp\left\{\ii S^{\rm eff}\left(A_{1},A_{2}\ldots,A_{N}\right)\right\}
\end{split}
\end{align}
where $A_i(s)$ corresponds to the gauge field of replica number $i$.
This defines an effective action describing the coupling of the gauge fields (of all replicas) to the Wilson line. This action has the following perturbative expansion
\begin{equation}
\ii S^{\rm eff}\left(A_{1},A_{2}\ldots,A_{N}\right) = \sum_{K=1}^{\infty} (\ii g_s)^K G_K
\label{eq:Seff}
\end{equation}
where the expressions for $\{G_K\}$ are straightforward to compute. The first three orders are given by  
\begin{subequations}
\label{Gdefs}
\allowdisplaybreaks
\begin{align}
G_1&=\int_{s_0}^{s_f} ds \sum_i A_i(s)\\
G_2&=\frac12 \int_{s_0}^{s_f} ds_1 \int_{s_0}^{s_f} ds_2\,
\Theta(s_2-s_1)\, \sum_i \left[A_i(s_2),A_i(s_1)\right]
+\frac12 \int_{s_0}^{s_f} ds \int_{s_0}^{s_f} dt\,
\sum_{i<j}\left[A_i(s),A_j(t)\right]
\,\\
\begin{split}
G_3&=\frac{1}{6} \int_{s_0}^{s_f} ds_1 \int_{s_0}^{s_f} ds_2\,\int_{s_0}^{s_f} ds_3\,\Theta(s_3-s_2)\,\Theta(s_2-s_1)\times\\&\times  
\sum_i \Big(  \left[\left[A_i(s_3),A_i(s_2)\right],A_i(s_1)\right]
- \left[\left[A_i(s_2),A_i(s_1)\right],A_i(s_3)\right]
\Big)\\
&+ \frac14 \int_{s_0}^{s_f} ds_1 \int_{s_0}^{s_f} ds_2\,\int_{s_0}^{s_f} dt \,\,
\Theta(s_2-s_1) \sum_{i\neq j} (-1)^{\Theta(j<i)} \left[\left[A_i(s_2),A_i(s_1)\right],A_j(t)\right]
\\
&+\frac{1}{12}\int_{s_0}^{s_f} ds \int_{s_0}^{s_f} dt\,\int_{s_0}^{s_f} du \sum_{i\neq j}\left[\left[A_i(s),A_j(t)\right],A_j(u)\right]\,+\,\frac{1}{6}
\int_{s_0}^{s_f} ds \int_{s_0}^{s_f} dt \,\int_{s_0}^{s_f} du\,\times\\&\times  
\sum_{i<j<k} \Big( \left[\left[A_i(s),A_j(t)\right],A_k(u)\right]
-  \left[\left[A_j(t),A_k(u)\right],A_i(s)\right]
\Big).
\end{split}
\end{align}
\end{subequations}\\
Importantly, because of the origin of these expressions in the BCH formula via eqs.~(\ref{repeated_BCH}) and (\ref{Path_ordered_repeated_BCH}) the general form of $G_K$ at any order can be simply expressed by relabelling integration variables, to give:
\begin{equation}
\label{G_K_def}
G_K=
Y_{i_{1}\ldots i_{n_{1}}} \circ
Y_{i_{n_{1}+1}\ldots i_{n_{1}+n_{2}}}\circ
\ldots \circ
Y_{i_{K-n_{b}+1}\ldots i_{K}}
\left(\prod_{i=1}^K\int ds_i\right)\,\,f_{i_1\,i_2\ldots i_K}(\{s_i\})\,,
\end{equation}
where the $K$ fields are grouped into $b$ sets such that $n_1+n_2+\ldots+n_b=K$.
Each of the $Y$-terms corresponds to a fully nested commutator of the form
\begin{equation}
\label{Ydef}
Y_{i_{1}\,i_{2}\,i_3\ldots i_{n}}=A_{i_{1}}(s_{1}) \circ  A_{i_{2}}(s_{2}) \circ A_{i_{3}}(s_{3}) \circ \ldots \circ  A_{i_{n}}(s_{n})
\end{equation}
where in both eqs.~(\ref{G_K_def}) and (\ref{Ydef}) we used the notation
\begin{equation}
\label{circ}
A_{i_1}(s_1) \circ  A_{i_2}(s_2) \circ A_{i_3}(s_3) \circ \ldots \circ  A_{i_K}(s_K)\equiv 
\left[\left[\ldots\left[A_{i_1}(s_1),A_{i_2}(s_2)\right],A_{i_3}(s_3)\right],\ldots,
A_{i_K}(s_K)\right]\,.
\end{equation}
To understand the resulting colour structure, which will be important below, note that we can factor out the fields 
$A_{i}(s)=A_{i}^{a}(s) T^a$ in (\ref{circ}), leaving behind a fully nested sequence of generators 
which can be recursively defined as  
\begin{equation}
\label{circ_rec_def}
T^{a_1} \circ \ldots \circ T^{a_n}  =  [ T^{a_1} \circ \ldots \circ T^{a_{n-1}}, T^{a_n}]
\end{equation}
with the innermost term being simply a commutator:
\[
T^{a_1} \circ T^{a_2} =[T^{a_1} , T^{a_2}]\,.
\]
Eq.~(\ref{circ_rec_def}) clearly corresponds to a fully connected non-Abelian tree graph: by repeated use of 
the colour algebra $[T^{a},T^{b}]=\ii f^{abc} T^c$ it reduces to a product of $K-1$ structure constants describing the sequence of branchings, multiplying a single generator $T^c$ corresponding to a single attachment to the Wilson line. 
The double-nesting structure of equations (\ref{G_K_def}) and (\ref{Ydef}) will be further discussed below.\\

We have seen that each term $G_K$ in eq.~(\ref{eq:Seff}) consists of nested commutators of gauge fields associated with different positions along the Wilson line, where the latter are integrated over. Each such structure is accompanied by a function of both replica indices and position parameters. The generating functional for the replicated theory has the final form
\begin{align}
\label{Zrepdef3}
\begin{split}
Z_{\rm rep.}\left[\left\{J^{a \mu (i)}\right\}\right]=\int{\cal D}A_1\,{\cal D}A_2\ldots{\cal D}A_N \
\exp \left[{\ii\sum_{i=1}^NS[A^{(i)}_\mu]}\right]\ \prod_{l=1}^L \exp \left[{\ii S^{\rm eff}_l\left(A_{1},\ldots,A_{N}\right) }\right],
\end{split}
\end{align}
where $\ii S^{\rm eff}_l$ denotes an effective action (eq.~\eqref{eq:Seff}) associated with Wilson
line $l$. \\

Having determined the generating functional in the replicated theory we can again use 
eqs. (\ref{Zrepdef_rel}) and (\ref{ZN}) to directly determine the exponent in the original theory by forming Feynman diagrams in the replicated theory and taking their ${\cal O}(N^1)$ parts.
To this end we may now determine the effective vertices $V_K$ associated with the emission of $K$ gluons from a given line~$l$.  One may then proceed as usual, replacing the fields 
in $S^{\rm eff}\left(A_{1},A_{2}\ldots,A_{N}\right)$ of eq.~(\ref{eq:Seff}) by functional derivatives with respect to sources $J^{a \,\mu_s \,(m)}(s)$, which may then be taken out of the functional integral. Equivalently one may obtain the effective vertices via
\begin{align}
\label{Vn_derivatives}
\begin{split}
V_1^{\mu_s}(a,m,s) &=\frac{\delta {\left(\ii S^{\rm eff}_l\right)}}{\delta A^{a}_{\mu_s\, m}(s)}
\\
V_2^{\mu_s \, \mu_t}\left((a,m,s);(b,n,t)\right) 
&=\frac{\delta {\left(\ii S^{\rm eff}_l\right)}}{\delta A^{a}_{\mu_s\,  m}(s)\, \delta A^{b}_{\mu_t\,  n}(t)}
\\
V_3^{\mu_s \, \mu_t \, \mu_u }\left((a,m,s);(b,n,t);(c,p,u)\right)  &=\frac{\delta 
{\left(\ii S^{\rm eff}_l\right)}}{\delta A^{a}_{\mu_s\,m}(s)\, 
\delta A^{b}_{\mu_t\, n}(t) \, \delta A^{c}_{\mu_u p}(u)},
\end{split}
\end{align} 
denoting the gauge field by
$A^{a}_{\mu_s\, m}(s)$ where $m$ is the replica index, $a$ an adjoint colour index and $s$ the position along the line 
such that $\beta^{\mu_s}=\dot{x}^{\mu_s}(s)$ is the velocity tangent to the Wilson line\footnote{Usually one considers straight Wilson lines, however the current analysis applies also  to the case of general contours.} at position $s$.  
The effective vertices $V_K$ are versions of the $G_K$ with additional symmetry
from the structure of the derivatives. Comparing eq.~(\ref{Zrepdef3}) with
eq.~(\ref{Zrepdef2}), it is clear what the role of these additional vertices is:
for gluon emissions which have the same replica number, they subtract
contributions from exponentiated lower-order vertices in order to implement path
ordering, and for gluon emissions with
different replica numbers they implement replica ordering.\\

The general structure of the resulting effective vertices is the following sum
\begin{equation}
\label{V_K}
V_K^{(l)\, \mu_{s_1}\ldots \mu_{s_K}}\big(\{(a_i,m_i,s_i)\}\big)= (\ii g_s)^K \sum_{j=1}^{(K-1)!} C_{K,j}^{(l)}\,\left(\prod_{i=1}^K\int ds_i \beta^{\mu_{s_i}}\right)\,\,v_{K,j}\big(\{(m_i,s_i)\}\big)\,,
\end{equation}
where the subscript $K$ on $V_K^{(l)\, \mu_{s_1}\ldots \mu_{s_K}}$ indicates the type of vertex (i.e. the number of emitted gluons) while the superscript $(l)$ indicates the relevant Wilson line. 
The vertex $V_K^{(l)\, \mu_{s_1}\ldots \mu_{s_K}}$ describes the emission of $K$ gluons with the following attributes: adjoint colour indices $a_i$, replica indices $m_i$, Lorentz indices $\mu_{s_i}$ and positions $s_i$ along the Wilson line. In order to identify the various indices associated with a given gluon, we keep the position labels as arguments on the left-hand side of eq. (\ref{V_K}) despite the fact that these are integrated over.
The function $\,v_{K,j}\big(\{m_i\},\{s_i\}\big)\,$ collects the kinematic and replica-number dependence. By construction the latter involves restrictions of the replica number assignment (as can be seen already in eq.~(\ref{Gdefs})) that can be written in terms of Kronecker delta and Heaviside functions. 
The colour factors in turn, take the form of nested commutators as dictated by the field commutators in eqs. (\ref{G_K_def}) and (\ref{Ydef}), namely
\begin{equation}
\label{CKj}
C_{K,j}^{(l)}=B_l^{\alpha_{1}\ldots \alpha_{n_{1}}} \circ
B_l^{\alpha_{n_{1}+1}\ldots \alpha_{n_{1}+n_{2}}}\circ
\ldots \circ
B_l^{\alpha_{K-n_{b}+1}\ldots \alpha_{K}}
\end{equation}
with
\begin{equation}
\label{BT}
B_l^{\alpha_{1}\,\alpha_{2}\ldots \alpha_{n}}=
T_l^{\alpha_{1}}\circ  T_l^{\alpha_{2}} \circ \ldots \circ  T_l^{\alpha_{n}}\,,
\end{equation}
where we again used the notation of eq.~(\ref{circ_rec_def}) for nested commutators. 
We define $\alpha_i= a_{\pi(j)_i}$ where $\pi(j)$ is a particular permutation $j$ of the colour matrices $T_l^{a_i}$ on Wilson-line~$l$ (identifying the relevant colour representation).
Explicit expressions for the effective vertices $V_1^{(l) \,\mu_s }$,  $V_2^{(l) \,\mu_s \,\mu_t}$ and $V_3^{(l)\,\mu_s \,\mu_t\, \mu_u}$ will be given in section \ref{sec:vertices}, see eqs.~(\ref{eq:V1}), (\ref{eq:V2}) and (\ref{eq:V3}),  respectively. 
\\

\begin{figure}[htb]
\begin{center}
\scalebox{.9}{\includegraphics{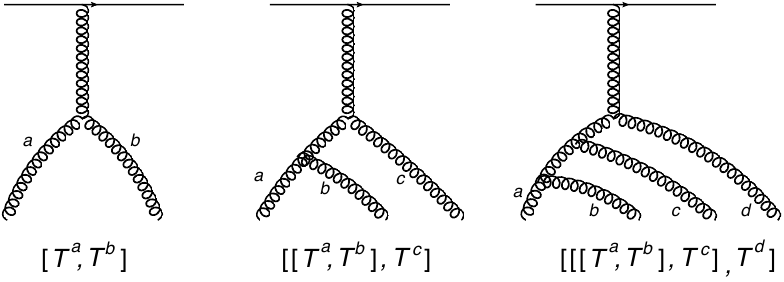}}
\\
\vspace*{20pt}
\scalebox{.9}{\includegraphics{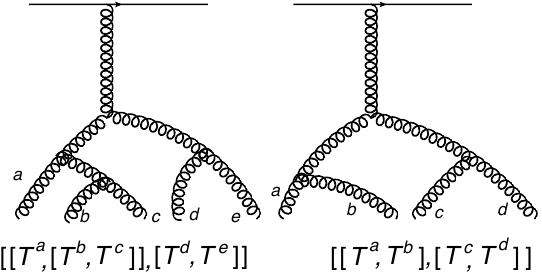}}
\caption{Examples of the tree colour structures associated with nested commutators: each corresponds to a connected colour factor. Here (a) illustrates the structures obtained from a single hierarchy of nested commutators as in eq.~(\ref{BT}) and (b) illustrates additional structures occurring in (\ref{CKj}) owing to the double hierarchy of nested commutators. }
\label{fig:vertices}
\end{center}
\end{figure}

At this point we wish to draw some general conclusions concerning the
effective-vertex based computation of the generating functional of eq.~(\ref{Zrepdef3}), with the goal of proving theorem \ref{theorem}.
We observe that the configuration-space structure of the effective vertices is non-local, consisting of all terms occurring in the function $v_{K,j}$.  
These functions become increasingly complicated at higher orders simply because there are many combinations of replica numbers which must be taken into account, each with a non-trivial number of permutations that must be subtracted to ensure path ordering. 
However the colour structure of the effective Feynman rules is very simple: as summarized by 
equations (\ref{CKj}) and (\ref{BT}) these are nested commutators, which by repeated use of 
the colour algebra, correspond to fully connected graphs such that one gluon is emitted from the Wilson line and then splits into $K$ gluons via a tree graph involving $K-1$ three-gluon vertices. 
A few examples with up to five gluon emissions are shown in figure~\ref{fig:vertices}.
We emphasize that the double hierarchy structure of nested commutators summarized by eqs. (\ref{CKj}) and (\ref{BT}) spans the space of all tree diagrams describing $1\to K$ gluon scattering, leading to $(K-1)!$ independent connected colour structures after using the Jacobi identity\footnote{The number of independent colour factors of the $V_K$ vertex can be deduced from the known result of ref.~\cite{DelDuca:1999rs} which considered the equivalent problem of $K+1$ gluon scattering by counting products of generators in the adjoint representation.
The same conclusion can also be reached in the context of web-mixing matrices by considering the case of a web with $K+1$ Wilson lines and $K$ single gluon exchanges, with one Wilson line having $K$ gluon attachments, each of which connects to one of the other Wilson lines, $W_{(1,1,1,\ldots, K)}$. The problem of counting the independent colour factors maps to the problem of computing the rank (or trace) of the relevant mixing matrix, which was shown to be $r=(K-1)!$ in ref.~\cite{Dukes:2013wa}. A generalization of this will be discussed in a forthcoming paper~\cite{Dukes:2013gea}.}.
\\

\begin{figure}
\begin{center}
\scalebox{.9}{\includegraphics{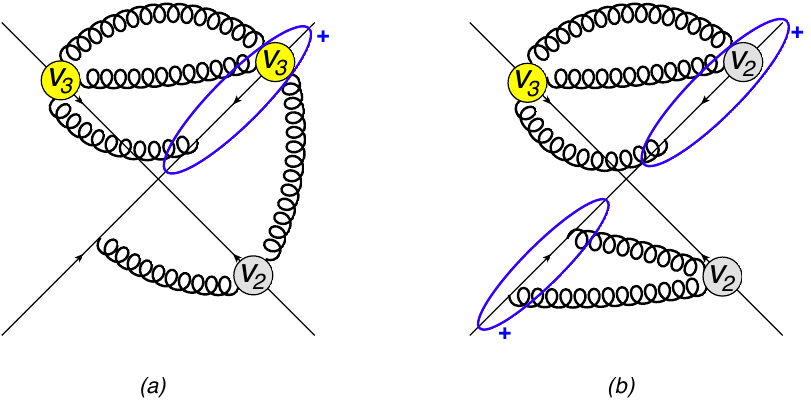}}
\caption{Examples of diagrams formed out of the effective Feynman rules appearing in eq.~(\ref{Zrepdef3}). 
For the multiple-emission effective vertices we indicated the type of vertex by $V_K$.
The colour part of each such vertex is internally fully connected. We also encircled all vertices on a given Wilson line by an ellipse and a ${\rm +}$ sign, to recall that one should symmetrise over all orderings of these, reflecting the ordinary (i.e. non-ordered) exponential in eq.~(\ref{Zrepdef3}).
}
\label{fig:diagex}
\vspace*{-10pt}
\end{center}
\end{figure}

Having established the structure of the effective vertices, let us now turn to
diagrams formed from these vertices in the replicated theory of eq.~(\ref{Zrepdef3}). We will see that the colour-connected nature of the former feeds directly into the latter. A general diagram in the replicated theory will consist of a given set of effective vertices on each Wilson line describing gluon emissions from that line, together with a way of joining up the emitted gluons. Both the colour and the configuration-space structure of the diagram is then fully determined by the internal structure of the vertices and the way they are joined up.
Examples are shown in figure~\ref{fig:diagex}. \\

We are now in a position to complete the proof that all ECFs are connected.  Consider a general diagram $D$, built up from these effective vertices. We wish to analyse the ECF of this diagram and to this end we will consider its colour factor in the replicated theory, $C_N(D)$, where the generating functional equals $Z^N$ (after identifying sources). 
Assume, without loss of generality, that $D=\cup_{i=1}^{m} S_D^{(i)}$ consists of $m$ subdiagrams $S_D^{(i)}$ where $S_D^{(i)}$ (for any $i=1..m$) is a connected subdiagram\footnote{Note that `connected' refers to the effective vertices: in configuration space $S_D^{(i)}$ need not be connected, due to the non-local nature of the effective vertices themselves.}, consisting of a fully joined up collection of effective vertices (for example in figure~\ref{fig:diagex}, $(a)$ consists of a single subdiagram ($m=1$) while $(b)$ consists of two subdiagrams ($m=2$)). 
Given that the colour part of each effective vertex is fully connected, any subdiagram $S_D^{(i)}$ has a fully connected colour factor. 
Recall now that each vertex $V_K$ in this subdiagram involves a sum over $K$ replica indices, and depends on these only though Kronecker delta or Heaviside functions. Propagators and gluon interactions off the Wilson lines (which connect these effective vertices in $S_D^{(i)}$) only involve Kronecker delta functions, since different replicas are non-interacting.  
Consequently, upon summing over all possible assignments of the replica numbers, each connected subdiagram $S_D^{(i)}$ acquires polynomial dependence on the total number of replicas $N$, starting with a linear term:
\begin{equation}
C_N\left(S_D^{(i)}\right)\propto N^1+\ldots\,,
\end{equation}
where the dots stand for higher powers in $N$ (examples of this may be seen in sections~\ref{sec:vertices} and~\ref{sec:effect-vert-pract}). 
Consider now the colour factor of the full diagram $D=\cup_{i=1}^{m} S_D^{(i)}$.
Because the assignments of replica numbers in each of the subdiagrams $S_D^{(i)}$ are mutually independent, it follows that  
\begin{equation}
C_N(D)\propto \prod_{i=1}^m C(S_D^{(i)}) \propto N^m+\ldots\,.
\end{equation}
Namely, the lowest order term in the polynomial is ${\cal O}(N^m)$. Referring to figure~\ref{fig:diagex}, for example, (a) and (b) have colour factors whose expansion in powers of $N$ start at ${\cal O}(N)$ and ${\cal O}(N^2)$ respectively. We thus see that the only diagrams which are ${\cal O}(N)$ have only single connected pieces, such that they have fully connected colour factors. By the replica trick argument, we arrive at the final result, which we set out to prove: {\it all exponentiated colour factors are fully connected}.\\

Some further comments are in order. Firstly, it is hopefully now clear why we removed explicit path and replica ordering in eq.~(\ref{Zrepdef3}): in this representation for the replicated generating functional, different replica numbers become completely decoupled from each other, with replica ordering implemented by higher order effective vertices. It is this fact that allows us to state that the colour factor for a graph with $m$ connected pieces (in the effective vertex language) is ${\cal O}(N^m)$. This decoupling is not manifest in the expression of eq.~(\ref{Zrepdef2}) -- there different replica numbers interact due to the ${\cal R}$ operator, such that in graphs with more than one connected piece, these pieces mix with each other. It should be stressed that the meaning of ``connected subdiagram'' is not the same in the languages defined by eq.~(\ref{Zrepdef2}) and eq.~(\ref{Zrepdef3}) respectively. In the latter case, we mean subdiagrams that may contain  effective vertices, which act (in part) to implement the replica ordering which is generated by ${\cal R}$ in the former approach. \\

\subsection{Effective vertices}
\label{sec:vertices}

\begin{figure}[b]
\begin{center}
\scalebox{.6}{\includegraphics{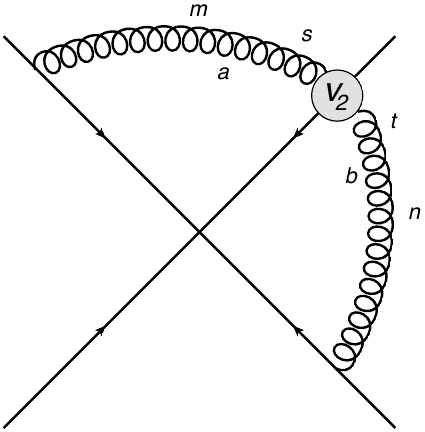}}
\caption{An effective-vertex connected graph representing the 1-2-1 web of fig.~\ref{fig:121}.  
}
\label{fig:121_vertex}
\end{center}
\end{figure}
In this section we give the explicit expressions for $V_1^{(l) \,\mu_s }$,  $V_2^{(l) \,\mu_s \,\mu_t}$ and $V_3^{(l)\,\mu_s \,\mu_t\, \mu_u}$  and provide simple examples to clarify their use. We postpone more elaborate examples to section~\ref{sec:effect-vert-pract}. \\

Using eqs.~(\ref{Vn_derivatives}) and
suppressing the $(l)$ labels for brevity we find the following first two vertices:
\begin{subequations}
\begin{align}
  \label{eq:V1}
&V_1^{\mu_s}\left(a,m,s\right) = \ii g_s T^a \int ds \beta^{\mu_s} , \\
\begin{split}
  \label{eq:V2}
&V_2^{\mu_s\,\mu_t}\left((a,m,s);(b,n,t)\right) 
\\&
= \frac{-\ii g_s^2}2 f^{ab c}T^c \int ds \beta^{\mu_s} \, dt  \beta^{\mu_t}
 \left[\Big( \Theta(s>t) - \Theta(t>s) \Big) \delta_{mn} +
  \Big( \Theta(m<n)-\Theta(n<m) \Big)\right]\,,
\end{split}
\end{align}
\end{subequations}
where we have chosen to write the commutator 
$T^{a}\circ  T^{b}$ directly as $\ii f^{a b c} T^{c}$. 
The $\Theta$-functions over the positions pick out individual (or subsets) of
diagrams in the web.  After contraction with the rest of the diagram, there is a
final sum over the replica numbers $m$ and $n$ and the $\mathcal{O}(N)$
coefficient is extracted to give the final result.  $V_1$ is unchanged from the
conventional approach.\\

The simplest web which includes the new $V_2$-vertex is the 1-2-1 web discussed
in section \ref{sec:ex} and shown in fig.~\ref{fig:121}.  
In terms of the effective vertices there is only one connected graph we can
form, the one shown in figure~\ref{fig:121_vertex}, which also shows the labels we
will use below.\\

The replica sum for the first term in eq.~\eqref{eq:V2} is
$\sum_{m,n=1}^N\delta_{mn}=N$.  Comparing to fig.~\ref{fig:121}, the condition
$\Theta(s>t)$ corresponds to diagram $A$ and $\Theta(t>s)$ corresponds to
diagram $B$.  The first term in eq.~\eqref{eq:V2} therefore gives
\begin{align}
  \label{eq:121frule}
  \frac{\ii }2 f^{abc}T^a_1T^c_2T^b_3\ N\
  \left(\mathcal{F}(A)-\mathcal{F}(B)\right),
\end{align}
where the ${\cal F}(X)$ terms indicate the kinematic factors of diagram $X$ and
we use the convention that these contain a factor of $(\ii g_s)^\ell$, where
$\ell$ is the number of gluon attachments to Wilson lines (here, $\ell=4$). The
second term gives zero when performing the replica sum and we immediately have
the results of eqs.~\eqref{R121}--\eqref{eq:finalctilde}.  \\

We now turn to the effective vertex for the emission of three gluons from a
single Wilson line, which is the first case where there is more than one independent colour structure.  If these gluons have adjoint indices $a$, $b$ and
$c$ respectively, then the three-fold nested commutators give terms with
the following colour factors:
\begin{align}
  \label{eq:jacobi}
\begin{split}
  C_{3,1}=\,&T^a\circ T^b\circ T^c=-f^{abd} f^{dce}T^e, \\
  C_{3,2}=\,&T^a\circ T^c\circ T^b=-f^{acd} f^{dbe}T^e,\\
  &T^b\circ T^c\circ T^a=-f^{bcd} f^{dae}T^e.
\end{split}
\end{align}
These are related by the Jacobi Identity and hence are not independent. We therefore choose to express the third in terms of the first two. While this choice is necessary for a linearly- independent colour basis, it does remove the explicit three-fold symmetry of the vertex (clearly this symmetry still exists in the sum of terms). The resulting $V_3$ is:
\begin{align}
  \label{eq:V3}
\begin{split}
  V_3^{\mu_s \,\mu_t\, \mu_u}&\left((a,m,s);(b,n,t);(c,p,u)\right)= \ii g_s^3  \int ds  \beta^{\mu_s} \, dt  \beta^{\mu_t}\,  du \beta^{\mu_u}
\\&
  \times \,\left(f^{ab d} f^{dc e}T^e v_{3,1}((m,s),(n,t),(p,u)) +f^{ac d} f^{dbe}T^e v_{3,2}((m,s),(n,t),(p,u))\right),
\end{split}
\end{align}
where $v_{3,1}$ and $v_{3,2}$ carry the replica number and position attributes of the three gluons and thus ultimately correspond to linear combinations of kinematic factors of diagrams.  
They are given by
\begin{subequations}
\allowdisplaybreaks
\begin{align}
  \begin{split}
  \label{v31}
    v_{3,1} ((m,s),(n,t),(p,u))=\, &\Bigg\{ \frac16 \delta_{mnp} \Big(
    2\Theta(s>t>u) -\Theta(u>s>t) - \Theta(t>s>u) \\ & \qquad \qquad
    +2\Theta(u>t>s) - \Theta(t>u>s) -\Theta(s>u>t) \Big) 
\\    & \quad +\frac14 \delta_{mn} (-1)^{\Theta(p<n)}\cancel{\delta}_{np} \Big(
    \Theta(s>t) - \Theta(t>s) \Big)
\\    & \quad 
+\frac14 \delta_{np} (-1)^{\Theta(m<n)} \cancel{\delta}_{mn} \Big( \Theta(u>t)
     - \Theta(t>u) \Big)\\
    & \quad + \frac1{12} \Big( \delta_{np} + \delta_{mn}
- 2\delta_{mp}  \Big) \\
    & \quad +\frac16 \Big( 2\Theta(m<n<p) + 2\Theta(p<n<m) - \Theta(p<m<n) 
\\
    & \qquad \qquad
    -\Theta(n<m<p) - \Theta(n<p<m) - \Theta(m<p<n) \Big)\Bigg\}
  \end{split}
\\
\begin{split}
\label{v32}
    v_{3,2} ((m,s),(n,t),(p,u))=\,& \Bigg\{ \frac16 \delta_{mnp} \Big(
        2\Theta(s>u>t) -\Theta(t>s>u) - \Theta(u>s>t) \\ & \qquad \qquad
        +2\Theta(t>u>s) - \Theta(s>t>u) -\Theta(u>t>s) \Big) \\
        & \quad+\frac14 \delta_{mp} (-1)^{\Theta(n<p)}\cancel{\delta}_{np} \Big(
        \Theta(s>u) - \Theta(u>s) \Big)\\
        & \quad+\frac14 \delta_{np} (-1)^{\Theta(m<p)}\cancel{\delta}_{mp}  \Big(
        \Theta(t>u)- \Theta(u>t) \Big)\\
        & \quad + \frac1{12} \Big( \delta_{np} + \delta_{mp}
        - 2\delta_{mn}  \Big) \\
        & \quad +\frac16 \Big( 2\Theta(m<p<n) + 2\Theta(n<p<m) - \Theta(n<m<p) \\
        & \qquad \qquad
        -\Theta(m<n<p) - \Theta(p<m<n) - \Theta(p<n<m) \Big)\Bigg\}.
  \end{split}
\end{align}
\end{subequations}
Here there is no sum over repeated indices. We also introduced the notation $\cancel{\delta}_{np}$ which equals one for any $n\neq p$ and zero for $n=p$.\\

\begin{figure}[htb]
\begin{center}
\begin{minipage}[b]{0.6\linewidth}
\scalebox{.9}{\includegraphics{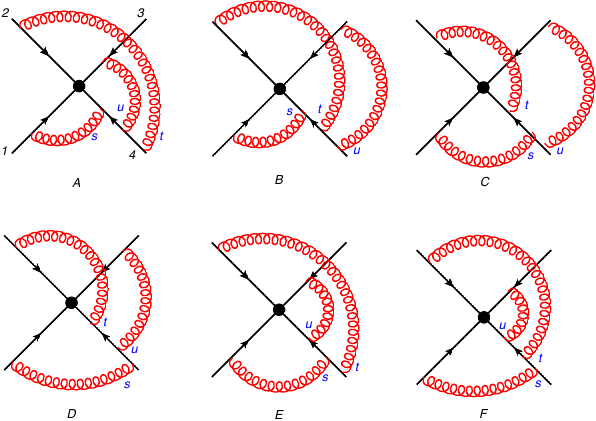}}
\end{minipage}
\begin{minipage}[b]{0.12\linewidth}
\begin{tabular}{c}
\scalebox{.41}{\includegraphics{large_arrow.pdf}}
\\
\\
\\
\\
\\
\\
\\
\\
\\
\\
\\
\\
\\
\\
\,
\end{tabular}
\end{minipage}
\begin{minipage}[b]{0.1\linewidth}
\scalebox{0.8}{\includegraphics{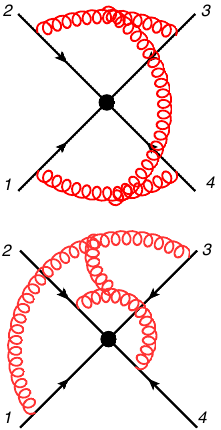}}
\begin{tabular}{c}
\\
\end{tabular}
\end{minipage}
\vspace*{-90pt}

\caption{The (1,1,1,3) web, together with a representation for the corresponding
  connected colour factors. Note that in each of the web diagrams $s$, $t$ and $u$ indicate respectively the positions along line 4 of the attachments of the gluons connecting to lines $1$, $2$ and $3$.} 
\label{fig:1113sect2}
\end{center}
\end{figure}

\begin{figure}[htb]
\begin{center}
\vspace*{10pt}
\scalebox{.7}{\includegraphics{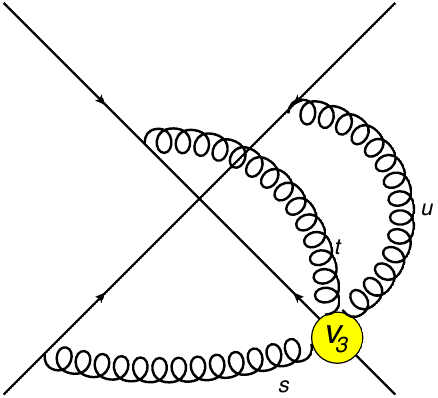}}
\caption{An effective-vertex connected graph representing the 1-1-1-3 web of fig.~\ref{fig:1113sect2}. Note that  $s$, $t$ and $u$ have the same meaning as in fig.~\ref{fig:1113sect2}, only here these variables are internal to the $V_3$ vertex. 
}
\label{fig:web1113_vertices}
\end{center}
\end{figure}

The simplest example including this vertex is the 1-1-1-3 web shown in
figure~\ref{fig:1113sect2}. Again there is only one possible connected graph using the effective vertices, the one shown in figure~\ref{fig:web1113_vertices}.
 In this example, only the first two lines in each of the two expressions $v_{3,1}$ and $v_{3,2}$, the ones proportional to $\delta_{mnp}$, yield non-zero contributions when the replica sum is performed. In these expressions each Heaviside function can be readily identified as one of the diagrams in fig.~\ref{fig:web1113_vertices}, and one arrives at the following contribution to the exponent:
\begin{align}
  \label{eq:1113frule}
  \begin{split}
    W_{(1,1,1,3)} &= -\frac{1}6 f^{abe}f^{cde}T_1^a T_2^b
    T_3^c T_4^d \,\,\Big(-{\cal F}(A) +2{\cal F}(B) -{\cal F}(C) -{\cal
      F}(D) -{\cal F}(E) +2{\cal F}(F)\Big)  \\
    & \quad - \frac{1}6 f^{ace}f^{bde}T_1^a T_2^b
    T_3^c T_4^d\,\,\Big(2{\cal F}(A) -{\cal F}(B) -{\cal F}(C) +2{\cal
      F}(D) -{\cal F}(E) -{\cal F}(F)\Big).
\end{split}
\end{align}
Here an important comment is due concerning the choice of colour basis. Recall that in writing $V_3$ we already made a choice to use the two colour factors $C_{3,1}$ and $C_{3,2}$ of eq.~(\ref {eq:jacobi}), eliminating the third via the Jacobi identity. This dictates the structure of the  1-1-1-3 web result in eq.~(\ref{eq:1113frule}). Equally well we could have chosen any pair, and different choices will be presented in Table \ref{tab:4leg3loop} below, where a global colour basis for all three-loop diagrams is used, and in appendix \ref{sec:w1113}, where the basis elements are obtained as eigenvectors of the web mixing matrix. All these choices are equivalent and can be shown to be equal upon using the Jacobi identity. \\

In section~\ref{sec:effect-vert-pract} we will elaborate on the process of calculating webs with
these new vertices in more complicated examples. Before that let us discuss in more general terms the colour factors of connected diagrams made out of these vertices and relate it to the language of the mixing matrices of refs.~\cite{Gardi:2010rn,Gardi:2011wa}. \\

\subsection{The vertex basis for web colour factors}
\label{sec:basis}

We have seen in the previous subsection that contributions to the exponent are made of connected graphs which constitute a certain number $n_l$
of effective vertices $V^{(l)}_{K_{i}}$ on each Wilson line $l$ (here $i$ enumerates the vertices occurring in a given connected graph on this Wilson line, while $K_i$ identifies the vertex type; note that we suppress Lorentz indices for brevity). 
A given vertex has $(K_{i}-1)!$ colour structures (after the use of the Jacobi identity), for example $V_3$ in eq.~(\ref{eq:V3}) has two. Below we shall use the index $j_i$ to enumerate a particular colour structure. While these are a priori all independent -- which is realised for example in the case of the 1-1-1-3 web in eq.~(\ref{eq:1113frule}) -- we will see examples where contraction with the rest of the graph renders them mutually dependent.\\

Because the generating functional, eq.~(\ref{Zrepdef3}), involves an ordinary exponential rather than an ordered one, the generated diagrams involve a fully symmetric sum. Thus the contribution of the Wilson lines to a given connected graph with $n_l$ effective vertices on each Wilson line $l$ takes the form (again suppressing Lorentz indices):
\begin{equation}
\label{symm}
\prod_{l=1}^{L} \Big\{V_{K_1}^{(l)}V_{K_2}^{(l)}\ldots V^{(l)}_{K_{n_l}}\Big\}_{+}
\end{equation}
where
\begin{equation}
\Big\{V_1V_2\ldots V_n\Big\}_{+}=  \frac{1}{n!}\,\sum_{\pi\in S_n} V_{\pi_1} V_{\pi_2}\ldots V_{\pi_n}\,.
\end{equation}
It follows that the colour structure for this graph is a sum of terms of the form:
\begin{equation}
\label{colour_basis_element}
c_j^{(L)}=\prod_{l=1}^{L} \Big\{C_{K_1,j_1}^{(l)}\, C_{K_2,j_2}^{(l)}\ldots C_{K_{n_l},j_{n_l}}^{(l)}\Big\}_{+}
\end{equation}
where $j$ enumerates the possible sets $((K_1,j_1),(K_2,j_2),\ldots (K_{n_l},j_{n_l}))$ on each of the Wilson lines.
Different connected graphs made out of the different combinations of effective vertices are necessarily independent, thus providing a natural way to construct a basis. Furthermore this basis is complete: it must span the colour space since the exponent as a whole is constructed out of such connected graphs. \\

One can construct such basis elements considering any given web. 
In the language of the web mixing matrices $c_j^{(L)}$ corresponds to a particular eigenvector with eigenvalue 1. Identifying $c_j^{(L)}$ as $\sum_k Y_{jk}C_k$, where $C=[C_1,\ldots, C_d]$ is the vector of colour factors of the diagrams in the web, as in eq.~(\ref{Wform2}), we see that the basis element corresponds to a row in the diagonalizing matrix~$Y$. Upon using equations (\ref{CKj}) and (\ref{BT}) to substitute a relevant term in $C_{K,j}^{(l)}$ for a nested commutator we can immediately identify the linear combination of diagrams corresponding to $c_j^{(L)}$, and thus read off the entries of the relevant row in $Y$ (examples are provided in the appendix, see e.g. section~\ref{sec:w222}). By construction each set of ordered generations on each line can appear an integer number of times, and it thus follows that the relevant row in $Y$ contains only integer numbers. These can be positive or negative (or zero) due to the commutator structure. In certain classes of webs these entries are in the set $\{0,+1,-1\}$ making the vertex basis remarkably simple.\\

For a given web there would usually be $r$ independent basis elements where $r$
is the rank of the mixing matrix (the exception is webs that include gluons that
are emitted and absorbed by the same Wilson line, where there is some
redundancy, as shown in the examples of sections~\ref{sec:w123se} and \ref{sec:w114se}). 
An interesting example is provided by the 2-2-2 web of Section~\ref{sec:w222} where all four colour factors that are needed to span the colour space of graphs connecting three legs at three loops arise from the four different connected graphs formed as combinations of $V_1$ and $V_2$.
In general one may use the vertices to construct the complete basis for all webs at a given order, as is done for example in Section~\ref{sec:3loop} at three-loop order. \\

In this section we have proved that all the colour factors that appear in the
exponent are fully connected. This required the introduction of a new formalism where multiple gluon emission from a given Wilson line is represented by effective vertices. 
In the next section we proceed to illustrate the use of this new formalism in calculations. In the following section and in the appendix we explicitly give results for all the webs that connect three and four Wilson lines at three-loop order.\\

\section{The effective vertices in practice}
\label{sec:effect-vert-pract}

In this section we illustrate the use of the effective
vertices derived in the previous section as a tool to compute webs, namely to determine the connected colour factors with the corresponding linear combination of integrals.
It should be emphasised at the outset that an algorithm to compute webs is
already available via the mixing matrix approach of
refs.~\cite{Gardi:2010rn,Gardi:2011wa}. The effective vertex approach provides
an alternative procedure leading to the same, or equivalent, final answer. The
main advantage of the vertex approach is that the basis of connected colour
factors is fixed in advance\footnote{This may be contrasted with the mixing
  matrices approach, where the colour basis is fixed only in the process of
  diagonalising the matrix. }. As explained in the previous section and
demonstrated further in section \ref{sec:3loop} and the appendix, this allows one to find the projection of all webs contributing at a given order on a given global colour basis.\\

In order to compute the (connected) colour factors and the corresponding
kinematic factors which multiply them for a given web, there is a simple
prescription:
\begin{enumerate}
\item Construct diagrams with attachments $V_1$, $V_2$, ... on each Wilson line in all
  possible ways provided that the gluons would form one connected piece if the
  Wilson lines are removed\footnote{Diagrams where this is not the case can only
  contribute at ${\cal O}(N^2)$ or higher in replica number as described in the
  previous section.}.
\item The colour factor for a Wilson line with more than one vertex attachment is the
  symmetric sum of all colour factors of those vertices, as in eq.~(\ref{symm}).  This will be
  illustrated in the following figures with a blue ring and a ``$+$'' label.
\item Divide by the symmetry factor for the diagram as for Feynman rules in any
  field theory.
\item Perform the sums over replica number, keeping only the ${\cal O}(N)$
  terms.
\item Identify the integrals over the positions of the gluon attachments along the Wilson lines with the corresponding diagrams in the original web to find the linear combinations
  of kinematic factors, $\mathcal{F}$.  We absorb a factor of $(\ii
  g_s)^{\ell}g_s^t$ into these factors, where $t$ is the number of three-gluon
  vertices and $\ell$ is the number of attachments to Wilson lines.  For
  diagrams with single-gluon exchanges only, $\ell$ is twice the number of
  loops.
\end{enumerate}

\begin{figure}[htb]
\begin{center}
\begin{tabular}{c}
  \scalebox{0.5}{\includegraphics{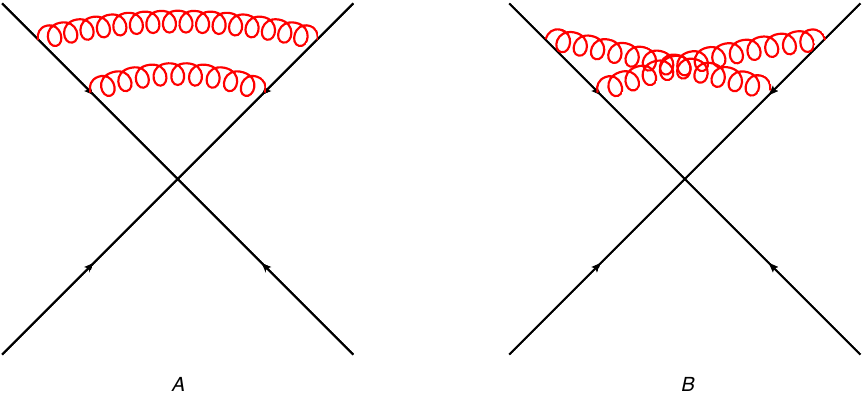}} \\
\\
  (a)\\
\\
  \scalebox{0.9}{\includegraphics{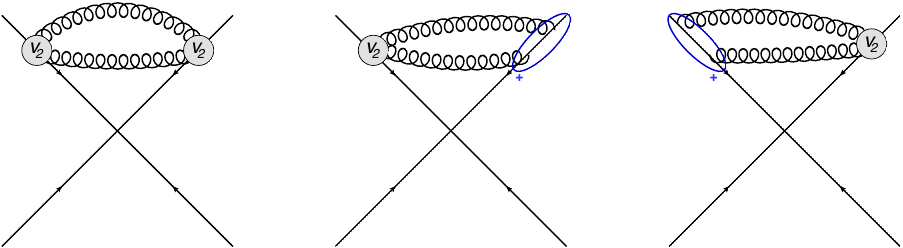}}\\
  (b)
\\
\end{tabular}
  \caption{(a) The two diagrams in $W_{(2,2)}$ and (b) the three effective-vertex diagrams that contribute.}
  \label{fig:W22}
\end{center}
\end{figure}

We begin with the simple example $W_{(2,2)}$. 
This web consists of two
diagrams, the ladder and the cross as shown in fig.~\ref{fig:W22}(a). 
This example is well known as it is relevant also in the context of a single Wilson loop studied in the 1980s~\cite{Gatheral:1983cz,Frenkel:1984pz}. Based on these results we know that only the cross diagram will contribute to the exponent. Let us then see how this result is reproduced in the present approach.\\

We begin with step (1) above and construct out of the effective vertices all
possible diagrams which are connected.  The three of these are shown in
fig.~\ref{fig:W22}(b) and involve only $V_1$ and $V_2$ of eqs.~(\ref{eq:V1})
and (\ref{eq:V2}).  Using the symmetric prescription for the diagrams which have
two single-gluon attachments ($V_1$s), the colour factor for the second diagram
is
\begin{align}
  \label{eq:W222ii}
  \ii f^{abc}T_1^c\ \times\ \frac12\{T_2^a,T_2^b\},
\end{align}
which is immediately zero.  The same is trivially true for the third diagram and
we conclude that the result for this web comes purely from the first diagram,
the one with a $V_2$ vertex on each line,  
\begin{align}
  \label{eq:W22i_first}
  \begin{split}
W_{(2,2)}(N)&=\frac12 \sum_{m,n=1}^N  V_2^{(1) \, \mu_{s_1}\,\mu_{t_1}}((a,m,s_1),(b,n,t_1))\,  
V_2^{(2)\, \mu_{s_2}\,\mu_{t_2}}((a,m,s_2),(b,n,t_2))\\
&
\times\, G^{(2,2)}_{\mu_{s_1}\,\mu_{t_1}\,\mu_{s_2}\,\mu_{t_2}} (s_1\beta_1, s_2\beta_2, t_1\beta_1,t_2\beta_2)  
  \end{split}
\end{align}
where we recall that the effective vertices $V_2$ contain integrals over the distance parameters and we define the combination of gluon propagators 
\[
G^{(2,2)}_{\mu_{s_1}\,\mu_{t_1}\,\mu_{s_2}\,\mu_{t_2}} (s_1\beta_1, s_2\beta_2, t_1\beta_1,t_2\beta_2)=\, D_{\mu_{s_1}\,\mu_{s_2}}(s_1\beta_1-s_2\beta_2)\,\, D_{\mu_{t_1}\,\mu_{t_2}}(t_1\beta_1-t_2\beta_2)\,,
\]
which forms part of the integrand.
The $\frac{1}{2}$ in eq.~(\ref{eq:W22i_first}) is a symmetry factor
corresponding to the interchange of the two gluons connecting the two vertices (step (3) above). 
Note that we already identified here the replica numbers between the two vertices, $m_1=m_2=m$ and $n_1=n_2=n$, since only identical replicas interact by gluon exchange. We have also identified the colour indices.
We denoted the expression in eq.~(\ref{eq:W22i_first}) by $W_{(2,2)}(N)$ to recall the fact that the right-hand side there depends on the number of replicas $N$: it is the result for this diagram in the replicated theory of eq.~(\ref{Zrepdef3}); to obtain the contribution to the exponent in the original theory, $W_{(2,2)}$, one needs to extract the coefficient of $N^1$ of this expression.\\

Upon using eq.~(\ref{eq:V2}) and substituting the expressions for $V_2$ 
into eq.~(\ref{eq:W22i_first}) we obtain: 
\begin{align}
  \label{eq:W22i}
  \begin{split}
    W_{(2,2)}(N)=\frac12\left(-\frac{\ii g_s^2}{2}\right)^2 &f^{abc}T_1^c\ f^{abd}T_2^d \sum_{m,n=1}^N 
 \int ds_1 \beta_1^{\mu_{s_1}} \, dt_1  \beta_1^{\mu_{t_1}}
 \int ds_2 \beta_2^{\mu_{s_2}} \, dt_2  \beta_2^{\mu_{t_2}}
\\&\times G^{(2,2)}_{\mu_{s_1}\,\mu_{t_1}\,\mu_{s_2}\,\mu_{t_2}} (s_1\beta_1, s_2\beta_2, t_1\beta_1,t_2\beta_2)\\
  &\times\Big( \left( \Theta(s_1>t_1) - \Theta(t_1>s_1) \right) \delta_{mn} +
    \left( \Theta(m<n)-\Theta(n<m) \right) \Big) \\
  &\times\Big( \left( \Theta(s_2>t_2) - \Theta(t_2>s_2) \right) \delta_{mn} +
    \left( \Theta(m<n)-\Theta(n<m) \right) \Big).
  \end{split}
\end{align}
We now perform the sum over the replica numbers $m$ and $n$ (step (4)) using 
\begin{equation}
\label{rep_sum}
\sum_{m,n=1}^N \delta_{nm}=N\,,\qquad\quad
\sum_{m,n=1}^N  \Theta(m<n)=\frac12 N(N-1)
\end{equation}
and find
\begin{align}
  \label{eq:replsum}
  \begin{split}
    W_{(2,2)}(N)&=-\frac{g_s^4}{8} f^{abc}T_1^c\ f^{abd}T_2^d 
\int ds_1 \beta_1^{\mu_{s_1}} \, dt_1  \beta_1^{\mu_{t_1}}
 \int ds_2 \beta_2^{\mu_{s_2}} \, dt_2  \beta_2^{\mu_{t_2}}
\\&\times G^{(2,2)}_{\mu_{s_1}\,\mu_{t_1}\,\mu_{s_2}\,\mu_{t_2}} (s_1\beta_1, s_2\beta_2, t_1\beta_1,t_2\beta_2)
\\ &\times \left[  N\Big( \Theta(s_1>t_1) - \Theta(t_1>s_1) \Big)\Big(
      \Theta(s_2>t_2) - \Theta(t_2>s_2) \Big) + N(N-1)\right]\,.
  \end{split}
\end{align}
We keep only the $\mathcal{O}(N^1)$ terms and translate the integrals into kinematic factors of the web diagrams of fig.~\ref{fig:W22}(a) (step (5)).  For the second term, in order to keep track of double-counting, we write
\[
\int ds_1 dt_1
ds_2 dt_2 \, 1 = \int ds_1 dt_1 ds_2 dt_2
\left( \Theta(s_1>t_1) + \Theta(t_1>s_1)\right)\left( \Theta(s_2>t_2) + \Theta(t_2>s_2)
\right)\,,
\] 
and we finally get the contribution of this web to the exponent
\begin{align}
  \label{eq:kintrans}
  \begin{split}
    W_{(2,2)}&=-\frac{1}{8}f^{abc}T_1^c\ f^{abd}T_2^d \bigg[ \Big(
      \mathcal{F}(A)-\mathcal{F}(B)-\mathcal{F}(B)+\mathcal{F}(A) \Big) -
    2\left( \mathcal{F}(A)+\mathcal{F}(B) \right) \bigg]\\
    &=\ \frac{1}{2}f^{abc}T_1^c\ f^{abd}T_2^d\ \mathcal{F}(B) = \frac{N_c}{2} T_1^c T_2^c \ \mathcal{F}(B).
  \end{split}
\end{align}
which is the expected result.
\\

\begin{figure}[hbt]
\centering
\begin{tabular}{c}
  \scalebox{0.6}{\includegraphics[angle=0]{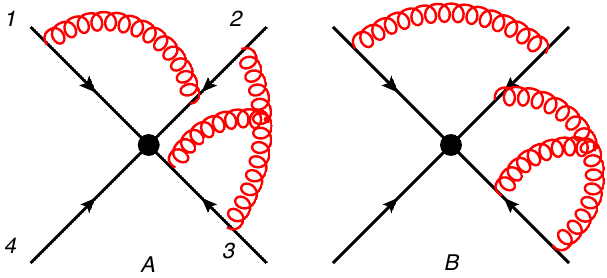}} \\
\\
(a)
\\
\\
  \scalebox{0.6}{\includegraphics{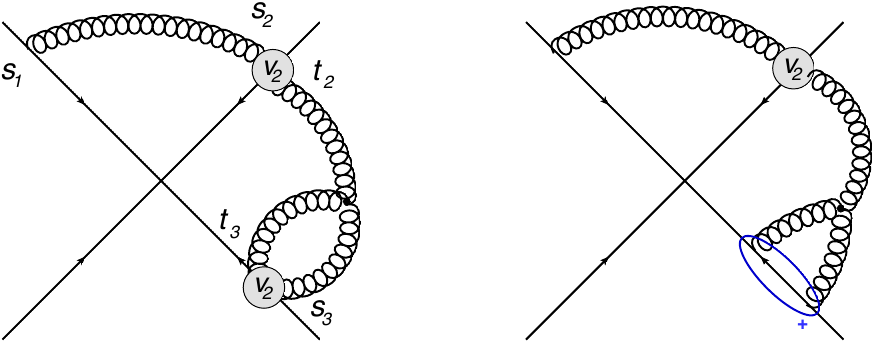}}\\
  (b)\\
\end{tabular}
  \caption{(a) The two diagrams in the $W^{3g}_{(1,2,2)}$ web. (b) The two
    diagrams which contribute to $W^{3g}_{(1,2,2)}$ in the vertex-approach.  The
    right-hand diagram is zero.}
  \label{fig:W122v}
\end{figure}

Further care is required with the ordering of gluons on a given line when the
vertices are used in combination with a three-gluon vertex.  This can be
illustrated by the 1-2-2 web, $W^{3g}_{(1,2,2)}$.  The mixing-matrix results for
this web are discussed in detail in Appendix~\ref{sec:w1223g}.  The web consists
of the two diagrams shown in fig.~\ref{fig:W122v}(a).  In the vertex approach,
there are also two possible diagrams which are depicted in
fig.~\ref{fig:W122v}(b).
The right-hand diagram is immediately zero due to the symmetric colour factor on
line 3 and the anti-symmetric three-gluon vertex.\\  

Thus, also here the complete answer for the web can be obtained from a single effective diagram, the one with a $V_2$ vertex on both line 2 and line 3, shown on the left-hand side in fig.~\ref{fig:W122v}. We find
\begin{align}
  \label{eq:v2twice}
  \begin{split}
W_{(1,2,2)}^{3g}(N)&=\frac12 \left(\ii g_s f^{bef}\right) 
\ \sum_{m,n,p,q}   \delta_{np}\delta_{pq}\,\,
 V_1^{(1)\, \mu_{s_1}}(a,m,s_1)\, V_2^{(2)\, \mu_{s_2}\, \mu_{t_2}}((a,m,s_2),(b,n,t_2))\\& \times \,V_2^{(3)\, \mu_{s_3}\, \mu_{t_3}}((e,p,s_3),(f,q,t_3))
\,\,  G^{{(1,2,2)}\,,{3g}}_{\mu_{s_1}\, \mu_{s_2}\, \mu_{t_2} \,\mu_{s_3}\, \mu_{t_3} }
(s_1\beta_1,s_2\beta_2,t_2\beta_2,s_3\beta_3,t_3\beta_3)
 \end{split}
\end{align}
where the integrand of the position integrals contains the function 
\begin{align}
  \label{eq:G122}
  \begin{split}
&G_{\mu_{s_1}\, \mu_{s_2}\, \mu_{t_2} \,\mu_{s_3}\, \mu_{t_3} }
(s_1\beta_1,s_2\beta_2,t_2\beta_2,s_3\beta_3,t_3\beta_3)
=\int d^{d}x 
\, \Gamma^{\nu\rho\sigma}(t_2\beta_2-x,\,t_3\beta_3-x,\,s_3\beta_3-x)
\\&\qquad\qquad \qquad\times 
D_{\mu_{s_1}\,\mu_{s_2}}(s_1\beta_1-s_2\beta_2)\,
\, D_{\mu_{t_2} \,\nu} (t_2\beta_2-x)
\, D_{\mu_{s_3} \,\rho}(s_3\beta_3-x)
\, D_{\mu_{t_3} \,\sigma}(t_3\beta_3-x)
 \end{split}
\end{align}
which collects the propagators and the kinematic part of the three-gluon vertex, $\Gamma^{\nu\rho\sigma}$.
The first factor of a half in (\ref{eq:v2twice}) is the symmetry factor and the extra
$\delta$-functions in the first line arise from the three-gluon vertex
which can only connect three gluons with the \emph{same} replica
number. Upon inserting the expressions for $V_1$ and $V_2$ of eqs.~(\ref{eq:V1})
and (\ref{eq:V2}) we get
\begin{align}
  \label{eq:v2twice2}
  \begin{split}
W_{(1,2,2)}^{3g}(N) &= \frac12 (\ii g_s) \left(-\frac{\ii g_s^2}2\right)^2 T_1^a\ f^{abc}T_2^c\ f^{efd}T_3^d\
  \left(\ii g_s f^{bef}\right)\ \sum_{m,n,p,q} \delta_{np}\delta_{pq} \\
  & \times \int ds_1 \beta_1^{\mu_{s_1}}
\, ds_2 \beta_2^{\mu_{s_2}} \, dt_2  \beta_2^{\mu_{t_2}}
\, ds_3 \beta_3^{\mu_{s_3}} \, dt_3  \beta_3^{\mu_{t_3}}
\, G_{\mu_{s_1}\, \mu_{s_2}\, \mu_{t_2} \,\mu_{s_3}\, \mu_{t_3} }
(s_1\beta_1,s_2\beta_2,t_2\beta_2,s_3\beta_3,t_3\beta_3)
\\&\times
  \Big( (\Theta(s_2>t_2)-\Theta(t_2>s_2) )\delta_{mn} + \Theta(m>n)-\Theta(n>m)
  \Big) \\
  &\times \Big( (\Theta(s_3>t_3)-\Theta(t_3>s_3) )\delta_{pq} +
    \Theta(p>q)-\Theta(q>p) \Big). \\
  \end{split}
\end{align}
We note that while the Heaviside functions allow both orders of $s_3$
and $t_3$, diagrams with $s_3>t_3$ contribute with a relative minus
sign compared to those with $s_3<t_3$ because, in order to have a
planar diagram like the orignal two, one changes the order of the
colour indices from $f^{bef}$ to $f^{bfe}$ in the
three-gluon vertex.\\

When one performs the sum over replica numbers, the only non-zero terms come from the terms proportional to $\delta_{mn}$ and $\delta_{pq}$ in the last two lines of eq.~(\ref{eq:v2twice2}).  For these terms, the sum is just $N$ and we find:
\begin{align}
  \label{eq:w122vres_N}
  \begin{split}
    W_{(1,2,2)}^{3g}(N) &= -\frac\ii8 f^{abc} f^{efd}
    f^{bef} T_1^a T_2^c T_3^d\ N\Big( {\cal F}(B) - (-{\cal
      F}(B)) -  {\cal F}(A) + (-{\cal F}(A)) \Big)\,.
 \end{split}
\end{align}
Finally, extracting the coefficient of $N^1$ we get the contribution of this web to the exponent:
\begin{align}
  \label{eq:w122vres}
  \begin{split}
        W_{(1,2,2)}^{3g}&= \frac{\ii}4 N_c\ f^{acd} T_1^a T_2^c T_3^d\ ({\cal
      F}(B) - {\cal F}(A)).
  \end{split}
\end{align}
This agrees with the result from the web-mixing matrices, eqs.~(\ref{ecf_122}) and (\ref{f1_122}).\\

\begin{figure}[htb]
  \centering
  \scalebox{0.90}{\includegraphics{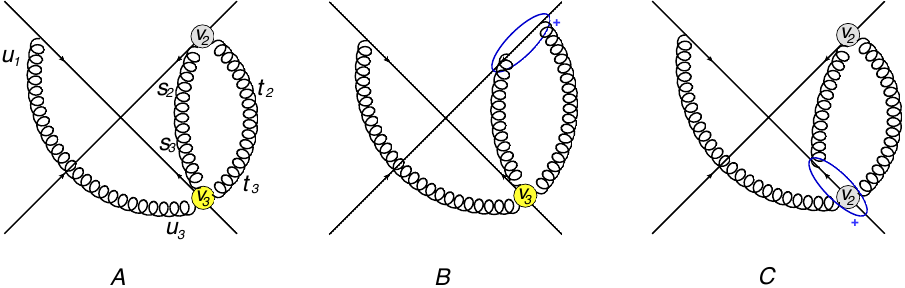}}\\
  \caption{The three diagrams which contribute to $W_{(1,2,3)}$ in the effective vertex approach.}
  \label{fig:w123eff}
\end{figure}

One can immediately see in both of these examples the advantage of quickly
returning a single connected colour factor with its corresponding kinematic
factor.  Our final example in this section is the 1-2-3 web, $W_{(1,2,3)}$.
This is treated in the web-mixing matrix approach in Appendix~\ref{sec:w123}.
Here there are three possible connected diagrams which can be constructed from
the effective vertices, as shown in fig.~\ref{fig:w123eff}.
 We find for the first diagram:
\begin{align}
  \label{eq:W123diag1}
  \begin{split}
W_{(1,2,3)}^A(N)&=\ \frac12\sum_{m,n,p} 
 V_1^{(1)\,\mu_{u_1}}(c,p,u_1)\, V_2^{(2)\,\mu_{s_2}\,\mu_{t_2}}((a,m,s_2),(b,n,t_2))
\\&\times 
V_3^{(3)\, \mu_{s_3}\, \mu_{t_3}\, \mu_{u_3} }((a,m,s_3),(b,n,t_3),(c,p,u_3))
\\&\times  G_{\mu_{u_1}\,\mu_{s_2}\,\mu_{t_2}\,\mu_{s_3}\, \mu_{t_3}\, \mu_{u_3}}(u_1\beta_1,\,s_2\beta_2,\,t_2\beta_2,\, s_3\beta_3,\,t_3\beta_3,\,u_3\beta_3)
  \end{split}
\end{align}
where 
\begin{align*}
 G^{(1,2,3)}_{\mu_{u_1}\,\mu_{s_2}\,\mu_{t_2}\,\mu_{s_3}\, \mu_{t_3}\, \mu_{u_3}}&(u_1\beta_1,\,s_2\beta_2,\,t_2\beta_2,\, s_3\beta_3,\,t_3\beta_3,\,u_3\beta_3)=\\&\qquad 
D_{\mu_{u_1}\, \mu_{u_3}} (u_1\beta_1-u_3\beta_3)\,
D_{\mu_{s_2}\, \mu_{s_3}} (s_2\beta_2-s_3\beta_3)\,
D_{\mu_{t_2}\, \mu_{t_3}} (t_2\beta_2-t_3\beta_3)\,.
\end{align*}
Upon inserting the expressions for $V_2$ and $V_3$ of eqs.~(\ref{eq:V2}) and~(\ref{eq:V3}) we get
\begin{align}
  \label{eq:W123diag1_}
  \begin{split}
W_{(1,2,3)}^A(N)&=\ \frac12 (\ii g_s)T_1^c\times \left(-\frac{\ii g_s^2}2\right) f^{abd}T_2^d\times \ii
    g_s^3 T_3^e \sum_{m,n,p} 
\\&\qquad \times 
\int du_1 \beta_1^{\mu_{u_1}} \,
ds_2 
\beta_2^{\mu_{s_2}} \,
dt_2
\beta_2^{\mu_{t_2}} \,
ds_3 
\beta_3^{\mu_{s_3}} \,
dt_3
\beta_3^{\mu_{t_3}} \,
du_3
\beta_3^{\mu_{u_3}} 
\\&\qquad \times  \Big( \left( \Theta(s_2>t_2) - \Theta(t_2>s_2) \right) \delta_{mn}
      +\Theta(m<n)-\Theta(n<m) \Big) \\  
    & \qquad \times \left( f^{abf} f^{fce} v_{3,1}((s_3,m),(t_3,n),(u_3,p)) + f^{acf}f^{fbe} v_{3,2}((s_3,m),(t_3,n),(u_3,p)) \right)\\
&  \qquad \times G^{(1,2,3)}_{\mu_{u_1}\,\mu_{s_2}\,\mu_{t_2}\,\mu_{s_3}\, \mu_{t_3}\, \mu_{u_3}}(u_1\beta_1,\,s_2\beta_2,\,t_2\beta_2,\, s_3\beta_3,\,t_3\beta_3,\,u_3\beta_3)\,
.
  \end{split}
\end{align}
Let us now consider the colour factors.
A priori, the $v_{3,1}$ and $v_{3,2}$ terms would contribute to two different colour
factors.  However, in this diagram they are both contracted with $f^{abd}$ from
the $V_2$ which means that the second is $\frac12$ times the first.  We may
rewrite the equation above then as
\begin{align}
  \label{eq:w123effv2}
  \begin{split}
    W_{(1,2,3)}^A(N)=&\ \left( \frac{\ii g_s^6}4 \right)  N_c f^{dce} T_1^c T_2^d T_3^e  \sum_{m,n,p} 
\int du_1 \beta_1^{\mu_{u_1}} \,
ds_2 
\beta_2^{\mu_{s_2}} \,
dt_2
\beta_2^{\mu_{t_2}} \,
ds_3 
\beta_3^{\mu_{s_3}} \,
dt_3
\beta_3^{\mu_{t_3}} \,
du_3
\beta_3^{\mu_{u_3}}    
\\&\qquad \times \left( \left( \Theta(s_2>t_2)
        -\Theta(t_2>s_2) \right) \delta_{mn} +\Theta(m<n)-\Theta(n<m) \right) \\ 
    & \qquad \times \left( v_{3,1}((s_3,m),(t_3,n),(u_3,p)) +\frac12 v_{3,2}((s_3,m),(t_3,n),(u_3,p))\right)
\\
&\qquad \times G^{(1,2,3)}_{\mu_{u_1}\,\mu_{s_2}\,\mu_{t_2}\,\mu_{s_3}\,
  \mu_{t_3}\, \mu_{u_3}}(u_1\beta_1,\,s_2\beta_2,\,t_2\beta_2,\,
s_3\beta_3,\,t_3\beta_3,\,u_3\beta_3)\ .
  \end{split}  
\end{align}
It turns out that when following the usual steps of summing over replica numbers
and identifying diagrams from the kinematic integrals, the $v_{3,2}$ contribution
gives zero here.  Extracting the ${\cal O}(N^1)$ coefficient, the contribution for this diagram to the exponent is
\begin{align}
  \label{eq:finalw123}
  W_{(1,2,3)}^A= \frac{\ii}4 N_c f^{cde} T_1^c T_2^d T_3^e\times\left( -\mathcal{F}(B) + \mathcal{F}(C) \right).
\end{align}
We now turn to the second diagram in fig.~\ref{fig:w123eff}.  The colour factor for line 2 is now $\frac12
\{T_2^a,T_2^b\}$, which is symmetric in $a$ and $b$.  Therefore when it
contracts with the first term in the $V_3$, which contains $f^{abf}$, we get zero.
The $v_{3,2}$ contribution for this diagram is:
\begin{align}
  \label{eq:w123effd2}
\begin{split}
   W_{(1,2,3)}^B(N)&=\ \frac12  \times (\ii g_s)T_1^c\times (\ii g_s)^2 \frac12\{T_2^a,T_2^b\}\times \ii
    g_s^3  f^{acf} f^{fbe}  T_3^e 
    \sum_{m,n,p} 
\\& \times \, \int du_1 \beta_1^{\mu_{u_1}} \,
ds_2 
\beta_2^{\mu_{s_2}} \,
dt_2
\beta_2^{\mu_{t_2}} \,
ds_3 
\beta_3^{\mu_{s_3}} \,
dt_3
\beta_3^{\mu_{t_3}} \,
du_3
\beta_3^{\mu_{u_3}}    
 \,v_{3,2}((s_3,m),(t_3,n),(u_3,p)) \\&
\times \, \Big( \left( \Theta(s_2>t_2)
        -\Theta(t_2>s_2) \right) \delta_{mn} +\Theta(m<n)-\Theta(n<m) \Big)
 \\
& \times G^{(1,2,3)}_{\mu_{u_1}\,\mu_{s_2}\,\mu_{t_2}\,\mu_{s_3}\, \mu_{t_3}\, \mu_{u_3}}(u_1\beta_1,\,s_2\beta_2,\,t_2\beta_2,\, s_3\beta_3,\,t_3\beta_3,\,u_3\beta_3)\,.
\end{split}
\end{align}
Computing the replica number and kinematic factors as before then yields 
\begin{align}
  \label{eq:w123d2end}
   W_{(1,2,3)}^B=-\frac1{12} f^{acf} f^{fbe} T_1^c \{T_2^a,T_2^b\} T_3^e\, \Big(
  2\mathcal{F}(A)-\mathcal{F}(B) -
    \mathcal{F}(C) -\mathcal{F}(D) +2\mathcal{F}(E) -\mathcal{F}(F) \Big).
\end{align}
Finally we come to the third diagram in fig.~\ref{fig:w123eff}.  Again we
construct this piece by piece as 
\begin{align}
  \label{eq:w123d3}
  \begin{split}
  W_{(1,2,3)}^C(N)=&(\ii g_s) T_1^d\times \left(-\frac{\ii g_s^2}{2}\right) f^{ab c}T_2^c\times
  (\ii g_s) \, \left(-\frac{\ii g_s^2}{2}\right) \frac12 f^{bde} 
  \{T_3^e,T_3^a\}\, \sum_{m,n,p}\\
   & \qquad \times  
\int du_1 \beta_1^{\mu_{u_1}} \,
ds_2 
\beta_2^{\mu_{s_2}} \,
dt_2
\beta_2^{\mu_{t_2}} \,
ds_3 
\beta_3^{\mu_{s_3}} \,
dt_3
\beta_3^{\mu_{t_3}} \,
du_3
\beta_3^{\mu_{u_3}}    
\\&\qquad \times  
 \left( \Big( \Theta(s_2>t_2)
        -\Theta(t_2>s_2) \right) \delta_{mn} +\Theta(m<n)-\Theta(n<m) \Big) \\
    & \qquad \times \Big( \left( \Theta(t_3>u_3)
        -\Theta(u_3>t_3) \right) \delta_{np} +\Theta(n<p)-\Theta(p<n) \Big)
 \\
&\qquad \times G^{(1,2,3)}_{\mu_{u_1}\,\mu_{s_2}\,\mu_{t_2}\,\mu_{s_3}\, \mu_{t_3}\, \mu_{u_3}}(u_1\beta_1,\,s_2\beta_2,\,t_2\beta_2,\, s_3\beta_3,\,t_3\beta_3,\,u_3\beta_3)\,.
  \end{split}
\end{align}
Performing the replica sums and manipulating the kinematic factors yields the
final result
\begin{align}
  \label{eq:w123d3end}
   W_{(1,2,3)}^C=-\frac1{12}f^{ab c}f^{bde} T_1^d T_2^c \{ T_3^e,T_3^a\} \Big(
  4\mathcal{F}(A)+\mathcal{F}(B)+\mathcal{F}(C)+\mathcal{F}(D)-2\mathcal{F}(E)+\mathcal{F}(F)
  \Big).
\end{align}
We conclude that using the effective vertices has then given the final result for the contribution of this web to the exponent as a sum of three independent colour factors.  In this case, each diagram in fig.~\ref{fig:w123eff} has yielded a contribution to exactly one colour factor.
The result can then be read off as the sum of
eqs.~\eqref{eq:finalw123}, \eqref{eq:w123d2end} and \eqref{eq:w123d3end}, and as expected, it agrees with the mixing matrices based computation in section \ref{sec:w123} upon conversion to the same colour basis. 
\\

We have illustrated the use of the effective vertices in a few examples.  In the next section, we apply this method to classify all possible colour factors which arise at three loops and determine the relevant combination of integrals contributing to each one.

\section{Webs at three-loop order}
\label{sec:3loop}

In the previous sections we provided a proof that all exponentiated
colour factors are fully connected and developed an effective-vertex formalism to compute webs.
For applications, such as the
calculation of the soft anomalous dimension, it is useful to classify all the connected colour factors which appear. Here we undertake this task at three loops.  
In discussing the various webs that occur, it is useful to divide them
into distinct classes, depending on how many Wilson lines are linked by 
gluon exchanges. In what follows, we will not include those diagrams that
already consist of a fully connected graph (such as the one involving a four gluon vertex) -- these appear as single
diagrams in the exponent, whose ECF is the same as their conventional colour
factor. Furthermore, we do not include those webs which connect only two eikonal
lines, as these have already been intensively studied, beginning with the work
of refs.~\cite{Gatheral:1983cz,Frenkel:1984pz,Sterman:1981jc}.\\

Different webs are expected to contribute to the same colour factors, and furthermore, they are likely to give rise to similar integrals. 
It is therefore much more efficient to compute the anomalous dimension after having determined the relevant combinations in a given global colour basis. 
The combinations, which we list below, may be calculated from the web mixing matrices or by the
effective Feynman rules of the previous sections (we used both methods in most cases, as a check).  The explicit web mixing matrices for all these webs are given in appendix~\ref{sec:web-mixing-app}. As emphasised in section~\ref{sec:basis} an important advantage of the vertex formalism is that it allows one to pick a colour basis right from the start. This will be illustrated at three-loop order in what follows. 
\\

\subsection{Webs connecting four lines}
\label{sec:4lines}

We begin with webs which connect four Wilson lines, using again the   
$(n_1,n_2,\ldots n_L)$ notation discussed in section~\ref{sec:ex}. 
There are only two web topologies in this category involving three individual gluon exchanges, 
$W_{(1,1,2,2)}$ and $W_{(1,1,1,3)}$, and one containing a three-gluon vertex, $W_{(1,1,1,2)}$. 
In each case similar webs where the gluons connect different Wilson lines can be obtained by relabelling colour and kinematic indices (for example $W_{(1,1,2,2)}$ can be related to $W_{(1,2,2,1)}$). In the table below we include only distinct topologies.  \\

The three web topologies mentioned above contribute only two independent
colour factors.  We choose to take the following colour factors as our basis:
\begin{align}
  \label{eq:4legcolour}
  \begin{split}
     c_1^{(4)} &= -f^{ade} f^{bce} T_1^a T_2^b
    T_3^c T_4^d =T_1^{a}T_2^{b} [T_3^{b},T_3^{c}][T_4^{a},T_4^{c}]\\
    c_2^{(4)} &= -f^{abe} f^{cde}  T_1^a T_2^b T_3^c T_4^d=
    [T_1^{a},T_1^{b}]T_2^{b}T_3^{c}[T_4^{c},T_4^{a}],
  \end{split}
\end{align}
where we use the notation of eq.~(\ref{colour_basis_element}) where the superscript indicates the number of Wilson lines being connected.
There is a third colour factor in this set, $c_3^{ (4)}=-f^{ace} f^{bde}T_1^a T_2^b T_3^c T_4^d$; however, it can be written in terms
of the two above using the Jacobi Identity,
$c_3^{ (4)}=c_1^{ (4)}+c_2^{ (4)}$,
and hence it is not independent.  All of these could be mapped to a tree level soft gluon scattering topology, with end points on the Wilson lines. This is a consequence of the fact that four Wilson lines is the maximum number that can be connected by soft gluons at three loop order. For gluons connecting less than four Wilson lines, we will see both tree and loop level soft gluon topologies occuring in the connected colour factor graphs.\\

Full results for the mixing matrices and manipulation of colour factors for each
of the webs below appear in appendix~\ref{sec:web-mixing-app}.  The results are
summarised in the table below.  The kinematic coefficients are given here in terms of
linear combinations diagrams in that particular web.  The labels correspond to
the diagrams in the relevant appendix section.

\begin{table}[h]
  \centering
  \begin{tabular}{|c|c|c|c|}
    \hline
    Web Topology & Appendix & Colour Factor & Kinematic Coefficient \\ \hline
    $W_{(1,1,2,2)}$ & \ref{sec:w1122} & $c_1^{ (4)}$ & $\frac16(-2A-2B+C+D)$ \\ \hline
    $W_{(1,1,1,3)}$ & \ref{sec:w1113} & $c_1^{ (4)}$ & $\frac16(2A-B-C+2D-E-F)$
    \\  & & $c_2^{ (4)}$ & $\frac16(A+B-2C+D-2E+F)$ \\ \hline
    $W_{(1,1,1,2)}$ & \ref{sec:w1112} & $c_1^{ (4)}$ & $\frac12 (-A+B)$ \\ \hline
  \end{tabular}
  \caption{The three distinct topologies for three-loop webs connecting four
    Eikonal lines, and their decomposition in terms of the colour factors
    $c_1^{ (4)}$ and $c_2^{ (4)}$.  The letters in the kinematic coefficients
    refer to the kinematic parts of the labelled diagrams in the relevant appendix.   The ${\cal F}$s have been suppressed here for brevity.}
  \label{tab:4leg3loop}
\end{table}

\subsection{Webs connecting three lines}
\label{sec:3lines}

We may now repeat the same exercise for webs which connect three lines.  There
are more of these, with greater complexity as the number of Eikonal lines is no
longer maximal.  We must now include webs which involve a three-gluon vertex for
example.  The independent colour factors may be written in terms of four
independent ones in many ways.  We choose to use those suggested by the
effective vertices of section \ref{sec:proof} as follows:
\begin{align}
  \label{eq:3legcolour}
  \begin{split}
    c_1^{(3)} &= - f^{a c e} f^{bde} T_1^{\{a,b\}} T_2^c T_3^d
= \{T_1^{a},T_1^{b}\}[T_2^{b},T_2^{c}][T_3^{a},T_3^{c}]\\
    c_2^{(3)} &= - f^{cae} f^{bde} T_1^a T_2^{\{b,c\}}
    T_3^d 
= [T_1^{a},T_1^{b}]\{T_2^{b},T_2^{c}\}[T_3^{a},T_3^{c}]\\
    c_3^{(3)} &= - f^{cbe} f^{ade} T_1^a T_2^b
    T_3^{\{c,d\}}
= [T_1^{a},T_1^{b}][T_2^{b},T_2^{c}]\{T_3^{a},T_3^{c}\}\\
    c_4^{(3)} &= -\frac12 \ii f^{acd} f^{bef} f^{def} T_1^a T_2^b
    T_3^c = \frac12 \ii N_c f^{abc} T_1^a T_2^bT_3^c
= [T_1^{a},T_1^{b}][T_2^{b},T_2^{c}][T_3^{a},T_3^{c}].
  \end{split}
\end{align}

We classify the contribution from each web to a given colour factor in
table~\ref{tab:3leg3loop}, using the same conventions as the previous sub-section.\\
\begin{table}[h]
  \centering
  \begin{tabular}{|c|c|c|c|}
    \hline
    Web Topology & Appendix & Colour Factor & Kinematic Coefficient \\ \hline
    $W_{(1,2,3)}$ & \ref{sec:w123} & $c_2^{ (3)}$ & $\frac{1}{12} (2A-B-C-D+2E-F)$ \\
    && $c_3^{ (3)}$ & $-\frac{1}{12} (4A+B+C+D-2E+F)$ \\
    && $c_4^{ (3)}$ & $-\frac{1}{2} (B-C)$\\ 
    \hline
    $W_{(1,2,2)}^{3g}$ & \ref{sec:w1223g} & $c_4^{ (3)}$ & $-\frac{1}{2} (A-B)$ \\
    \hline
    $W_{(1,2,3)}^{\rm SE}$ & \ref{sec:w123se} & $c_4^{ (3)}$ & $-\frac12 (A-B)$ \\ \hline
    $W_{(1,1,3)}^{3g}$ & \ref{sec:w1133g} & $c_3^{ (3)}$ & $\frac12 A$ \\ 
    && $c_4^{ (3)}$& $-\frac{1}{2}(B-C)$ \\\hline
    $W_{(1,1,4)}^{\rm SE}$ & \ref{sec:w114se} & $c_3^{ (3)}$ & $\frac12 \big(
    A+B \big)$ \\
    && $c_4^{ (3)}$& $\frac12 (-A+B+C+D-E-F)$ \\ \hline
    $W_{(2,2,2)}$ & \ref{sec:w222} & $c_1^{ (3)}$ &
    $\frac{1}{12} (-2A-2B+C-2D+E-2F+G+H)$ \\
    && $c_2^{ (3)}$ & $\frac{1}{12}(-2A-2B-2C+D+E+F-2G+H) $ \\
    && $c_3^{ (3)}$ & $\frac{1}{12}(2A+2B-C-D+2E-F-G+2H) $ \\
    && $c_4^{ (3)}$ & $\frac12(-A+B) $ \\\hline
    $W_{(1,2,2)}^{3g\prime}$ & \ref{sec:w1223gp} & $c_2^{ (3)}$ & $-\frac{1}{4}(A-B+C-D)$ \\ 
    && $c_3^{ (3)}$ & $-\frac{1}{4}(-A+B+C-D)$ \\
    && $c_4^{ (3)}$ & $\frac12(A+B)$\\  \hline
  \end{tabular}
  \caption{The seven distinct topologies for three-loop webs connecting three
    Eikonal lines, and their decomposition in terms of the colour factors
    $c_i^{ (3)}$.  The letters in the kinematic coefficients
    refer to the kinematic parts of the labelled diagrams in the relevant
    appendix.  The ${\cal F}$s have been suppressed for brevity.}
  \label{tab:3leg3loop}
\end{table}

In order to compute the anomalous dimension at this order, it is necessary to
compute all possible permutations of the legs of these webs.  This process will
generate different colour factors for a given web to those which appear in the
table.  For example, $W_{(1,1,4)}^{\rm SE}$ will give contributions to
$c_1^{ (3)}$ and $c_2^{ (3)}$ when the leg with four attachments is labelled leg
1 or 2.  However, the tables in this section give all the necessary ingredients,
up to this relabelling.\\

The combined results of tables~\ref{tab:4leg3loop} and \ref{tab:3leg3loop}
therefore contain all the ingredients of the soft-gluon exponent at three-loop
order in terms of a basis of six independent colour factors and their
coefficients as linear combinations of kinematic factors of diagrams.  This
compact organisation of the calculation provides an example of the 
connected-colour-factor theorem as well as a first application of the effective-vertex formalism.

\section{Conclusions}
\label{sec:conclude}

The central result of this paper is that all exponentiated colour factors which
occur in the description of soft-gluon corrections to multi-parton scattering
correspond to \emph{fully-connected} graphs.  While the original non-Abelian exponentiation theorem refers to a Wilson loop~\cite{Gatheral:1983cz,Frenkel:1984pz}, which is directly relevant to the form factor, our proof now extends to the case of any number of Wilson lines, in arbitrary representations of the colour group.  
In the multi-parton case webs become closed sets of diagrams~\cite{Mitov:2010rp,Gardi:2010rn}   which may individually be reducible -- that is, involving gluon exchanges that are neither connected nor crossed -- 
and hence this constraint on their exponentiated colour factors is highly non-trivial.\\

This important result was proved using the replica trick, which was already used in refs.~\cite{Laenen:2008gt,Gardi:2010rn} to derive exponentiation properties. The latter reference made use of a replica-ordering operator, which in turn gave rise to web-mixing matrices. Here we proceeded by re-writing the operations of
replica-ordering and path-ordering through repeated use of the BCH formula as nested commutator terms in the exponent itself.  This leads to new effective vertices $V_K$ for the emission of $K$ gluons from a Wilson line, eq.~(\ref{V_K}).  These vertices embody the combinatorial complexity of replica and path ordering along the line through their non-local configuration-space and replica-index structure. Their colour factors, defined as nested commutators, are fully connected. The exponent as a whole is then  computed by forming fully connected graphs out of these vertices. \\

The effective-vertex based calculation, which has been demonstrated here in sections \ref{sec:vertices} and \ref{sec:effect-vert-pract}, is complementary to the mixing-matrix method of refs.~\cite{Gardi:2010rn,Gardi:2011wa,Gardi:2011yz,Dukes:2013wa}.   
Connected graphs made out of the effective vertices provide a convenient representation-independent colour basis for webs. Given that individual webs are not separately gauge invariant, fixing a basis is essential in order to combine webs with related colour factors so as to form ``gauge-invariant webs''. The availability of a basis becomes particularly important at the multi-loop level, where there is a priori much freedom in choosing a basis owing to Jacobi identities. The construction of gauge-invariant webs opens up the possibility of using innovative gauges to simplify the calculation of higher-loop soft-gluon effects (see e.g.~\cite{Chien:2011wz}).\\

Our finding that webs have connected colour factors adds a significant constraint on the structure of web mixing matrices. When put together with other properties of these matrices~\cite{Gardi:2010rn,Gardi:2011wa,Gardi:2011yz,Dukes:2013wa}, such as idempotence, this leads to further insights on the combinatorial structure of webs. One may go further than this and derive explicit solutions for web mixing matrices for any number of soft gluons in certain cases, or provide general formulae for the rank, corresponding to the number of independent colour factors. A first study in this direction will be presented in ref.~\cite{Dukes:2013gea}. \\

A pressing open problem is the calculation of the multi-parton soft anomalous
dimension at three loops, where diagrams that connect four Wilson lines can first be formed.  
We have demonstrated the connected-colour-factor result in this context by classifying all of the colour factors arising at three loops, and computing the linear combinations of kinematic factors which accompany each colour factor. The basis in this case contains two independent colour factors for webs connecting four Wilson lines, and four independent colour factors for webs connecting three lines. The calculation of the corresponding kinematic integrals is ongoing.

\vspace*{30pt}
{\bf \Large{Acknowledgements}}
\vspace*{10pt}

We are very grateful to Gregory Korchemsky for discussions at an earlier stage of this project, to Mark Dukes for ongoing collaboration on related topics, and to Claude Duhr for very helpful comments on the manuscript. EG and CDW are supported by the UK Science and Technology Facilities Council (STFC); JMS is funded by a Royal Society University Research Fellowship.
\vspace*{20pt}

\appendix

\section{Web mixing-matrix analysis for three-loop webs}
\label{sec:web-mixing-app}

In this section we collect the web mixing-matrices and matrices of left
eigenvectors for the three-loop webs considered in section~\ref{sec:3loop}.  As
in that section, we classify them according to how many Wilson lines they
connect.  We use the notation of section~\ref{sec:ex}, and give for each web:
the rank $r$, the mixing matrix $R$, the diagram definitions $A$, $B$,
$C\ldots$, the matrix of left eigenvectors $Y$ and its inverse.  We then
explicitly compute the colour factors $(YC)_i$ and the corresponding linear
combinations of kinematic factors, $f_i= ({\cal F}^TY^{-1})_i$, in each case.
We will use the shorthand notation
\[
T_i^{abcd...}=T_i^aT_i^bT_i^cT_i^d\ldots
\]
throughout. In addition, we will label each individual web diagram $D$ according to a notation first introduced in ref.~\cite{Gardi:2010rn}, as
\begin{equation}
\label{diag_notation}
D=\left[[s_1^{(1)},s_2^{(1)},\ldots s_{n_1}^{(1)}],\,\, [s_1^{(2)},s_2^{(2)},\ldots s_{n_2}^{(2)}],\,\, \cdots \,\, ,\,\,[s_1^{(L)},s_2^{(L)},\ldots s_{n_L}^{(L)}]\right]\,.
\end{equation}
Each of the square brackets corresponds to a different Wilson line (with $L$ in total). Each of the entries within a given bracket $[s_1^{(l)},s_2^{(l)},\ldots s_{n_l}^{(l)}]$ corresponding to leg $l$, is associated with an individual gluon attachment, where the order of the list indicates the order of gluon attachments to this leg: the list is ordered from the outside inwards toward the vertex (cusp).
The variables $s_i^{(l)}$ themselves are assigned values from 1 to $n_c$ (the number of connected pieces in~$D$) associating each of the gluons with a particular connected piece.

\subsection{Webs connecting four lines}
\label{sec:webs-connecting-four}
We begin by considering webs which connect four Wilson lines.  These results
are summarised in Section~\ref{sec:4lines}.

\subsubsection{$W_{(1,1,2,2)}$} 
\label{sec:w1122}
\begin{figure}[htb]
\begin{center}
\begin{minipage}[b]{0.7\linewidth}
\scalebox{.95}{\includegraphics{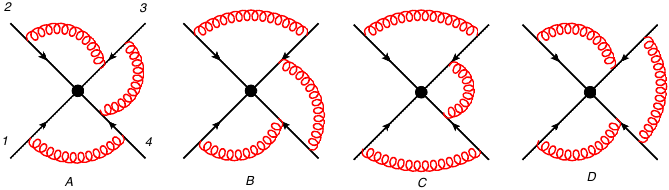}}
\end{minipage}
\begin{minipage}[b]{0.1\linewidth}
\begin{tabular}{c}
\scalebox{.4}{\includegraphics{large_arrow.pdf}}
\\
\\
\\
\\
\\
\,
\end{tabular}
\end{minipage}
\begin{minipage}[b]{0.1\linewidth}
\scalebox{1}{\includegraphics{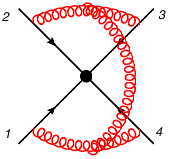}}
\end{minipage}

\vspace{-40pt}
\caption{The (1,1,2,2) web, together with the single connected colour factor one obtains.}
\label{fig:1122}
\end{center}
\end{figure}
\begin{align}
\begin{split}
&
r_{(1,1,2,2)}=1
\\&
R_{(1,1,2,2)}=\frac16
 \left[ \begin {array}{cccc} 2&2&-2&-2\\ \noalign{\medskip}2&2&-2&-2
\\ \noalign{\medskip}-1&-1&1&1\\ \noalign{\medskip}-1&-1&1&1
\end {array} \right] 
\,,\qquad
\left[ \begin {array}{c} A\\
\noalign{\medskip}B\\
\noalign{\medskip}C\\
\noalign{\medskip}D
\end{array}\right]
=
 \left[ \begin {array}{c} [[2],[1],[3,1],[2,3]]\\ \noalign{\medskip}[[
2],[1],[1,3],[3,2]]\\ \noalign{\medskip}[[2],[1],[1,3],[2,3]]
\\ \noalign{\medskip}[[2],[1],[3,1],[3,2]]\end {array} \right] 
\end{split}
\end{align}
\begin{align}
\begin{split}
&
Y_{(1,1,2,2)}= \left[ \begin {array}{cccc} -1&-1&1&1
\\ \noalign{\medskip}1/2&0&0&1
\\ \noalign{\medskip}1/2&0&1&0
\\ \noalign{\medskip}-1&1&0&0\end {array} \right] \,,\qquad
Y_{(1,1,2,2)}^{-1}= \left[ \begin {array}{cccc} -1/3&1/3&1/3&-1/3
\\ \noalign{\medskip}-1/3&1/3&1/3&2/3\\ \noalign{\medskip}1/6&-1/6&5/6
&1/6\\ \noalign{\medskip}1/6&5/6&-1/6&1/6\end {array} \right] 
\end{split}
\end{align}

For the single ECF we get:
\begin{align}
\begin{split}
(YC)_1&=T_1^cT_2^{a} \Big(-T_3^{ba}T_4^{cb}
-T_3^{ab}T_4^{bc}+T_3^{ab}T_4^{cb}+T_3^{ba}T_4^{bc}\Big)
\\
&=T_1^cT_2^{a} \Big(\ii f^{abd}T_3^dT_4^{cb}-\ii f^{abd}T_3^{d}T_4^{bc}\Big)
\\
&=T_1^cT_2^{a} T_3^dT_4^{e} f^{abd}f^{bce}
\end{split}
\end{align}
This colour factor corresponds to the diagram on the right hand side in figure \ref{fig:1122}. The corresponding kinematic factors for the $(1,1,2,2)$ web are
\begin{equation}
f_1=\frac16\Big(-2{\cal F}(A)-2{\cal F}(B)+{\cal F}(C)+{\cal F}(D)\Big).
\end{equation}

\subsubsection{$W_{(1,1,1,3)}$} 
\label{sec:w1113}

\begin{figure}[htb]
\begin{center}
\begin{minipage}[b]{0.6\linewidth}
\scalebox{.9}{\includegraphics{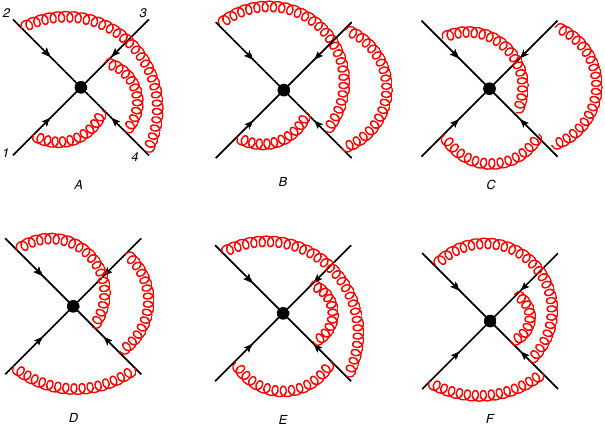}}
\end{minipage}
\begin{minipage}[b]{0.15\linewidth}
\begin{tabular}{c}
\scalebox{.5}{\includegraphics{large_arrow.pdf}}
\\
\\
\\
\\
\\
\\
\\
\\
\\
\\
\\
\\
\,
\end{tabular}
\end{minipage}
\begin{minipage}[b]{0.1\linewidth}
\begin{tabular}{c}
\scalebox{0.7}{\includegraphics{four_eikonals_1113_connected.pdf}}
\\
\\
\\
\\
\\
\\
\\
\\
\\
\\
\\
\\
\\
\end{tabular}
\end{minipage}
\vspace*{-150pt}

\caption{The six diagrams in the 1-1-1-3 web with the labels used below.  The
  right-hand side shows the two resulting colour factors.}
\label{fig:1113}
\end{center}
\end{figure}

\begin{align}
\begin{split}
&
r_{(1,1,1,3)}=2\\
&
R_{(1,1,1,3)}=\frac16
 \left[ \begin {array}{cccccc} 2&-1&-1&2&-1&-1\\ \noalign{\medskip}-1&
2&-1&-1&-1&2\\ \noalign{\medskip}-1&-1&2&-1&2&-1\\ \noalign{\medskip}2
&-1&-1&2&-1&-1\\ \noalign{\medskip}-1&-1&2&-1&2&-1
\\ \noalign{\medskip}-1&2&-1&-1&-1&2\end {array} \right] 
\,,\qquad
\left[ \begin {array}{c} A\\
\noalign{\medskip}B\\
\noalign{\medskip}C\\
\noalign{\medskip}D\\
\noalign{\medskip}E\\
\noalign{\medskip}F
\end{array}\right]
=
 \left[ \begin {array}{c} [[1],[2],[3],[2,3,1]]\\ \noalign{\medskip}[[
1],[2],[3],[3,2,1]]\\ \noalign{\medskip}[[1],[2],[3],[3,1,2]]
\\ \noalign{\medskip}[[1],[2],[3],[1,3,2]]\\ \noalign{\medskip}[[1],[2
],[3],[2,1,3]]\\ \noalign{\medskip}[[1],[2],[3],[1,2,3]]\end {array}
 \right] 
\end{split}
\end{align}
\begin{align}
\begin{split}
&
Y_{(1,1,1,3)}= \left[ \begin {array}{cccccc} -1&1&0&-1&0&1
\\ \noalign{\medskip}-1&0&1&-1&1&0\\ \noalign{\medskip}-1&0&0&1&0&0
\\ \noalign{\medskip}0&-1&0&0&0&1\\ \noalign{\medskip}1&1&0&0&1&0
\\ \noalign{\medskip}1&1&1&0&0&0\end {array} \right] \,,
\end{split}
\end{align}
\begin{align}
\begin{split}
&
Y_{(1,1,1,3)}^{-1}= \left[ \begin {array}{cccccc} -1/6&-1/6&-1/3&1/6&1/6&1/6
\\ \noalign{\medskip}1/3&-1/6&1/6&-1/3&1/6&1/6\\ \noalign{\medskip}-1/
6&1/3&1/6&1/6&-1/3&2/3\\ \noalign{\medskip}-1/6&-1/6&2/3&1/6&1/6&1/6
\\ \noalign{\medskip}-1/6&1/3&1/6&1/6&2/3&-1/3\\ \noalign{\medskip}1/3
&-1/6&1/6&2/3&1/6&1/6\end {array} \right] .
\end{split}
\end{align}
This leads to the following colour factors:
\begin{align}
\begin{split}
(YC)_1&=T_1^aT_2^bT_3^c\,\Big(-T_4^{bca}+T_4^{cba}-T_4^{acb}+T_4^{abc}\Big)=f^{bce}f^{ade}T_1^aT_2^bT_3^cT_4^d
\\
(YC)_2&=T_1^aT_2^bT_3^c\,\Big(-T_4^{bca}+T_4^{cab}-T_4^{acb}+T_4^{bac}\Big)=
f^{ace}f^{bde}T_1^aT_2^bT_3^cT_4^d 
\end{split}
\end{align}
and kinematic factors:
\begin{align}
\begin{split}
f_1&=\frac16\Big(-{\cal F}(A)+2{\cal F}(B)-{\cal F}(C)-{\cal F}(D)-{\cal F}(E)+2{\cal F}(F)\Big)
\\
f_2&=\frac16\Big(-{\cal F}(A)-{\cal F}(B)+2{\cal F}(C)-{\cal F}(D)+2{\cal F}(E)-{\cal F}(F)\Big)\,.
\end{split}
\end{align}

\begin{figure}
\begin{center}
\scalebox{.8}{\includegraphics{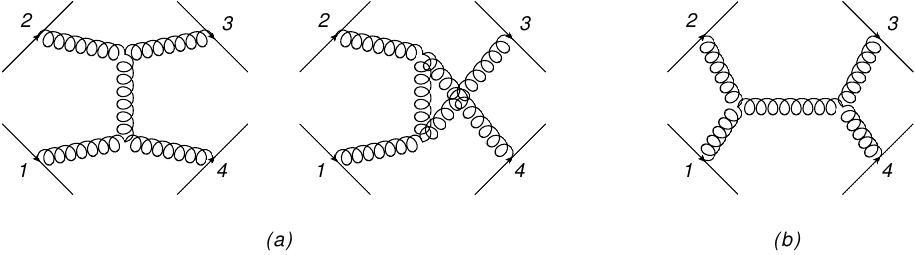}}
\caption{(a) Connected colour factors arising from the (1,1,1,3) web of figure~\ref{fig:1113}; (b) alternative connected colour factor, related to those of (a) by a Jacobi identity.}
\label{fig:1113b}
\end{center}
\end{figure}

Note that the choice of connected colour factors is not unique. Consider, for example, redrawing the right-hand side of figure~\ref{fig:1113} as shown in figure~\ref{fig:1113b}(a), where the Wilson lines are separated from each other. 
We see that the two connected colour factors of figure~\ref{fig:1113} correspond to tree-level soft gluon scattering topologies spanning the Wilson lines, and that the two connected colour factors correspond to the $t$- and $u$-channel topologies for the reaction $12\rightarrow 34$. However, the colour factor of the diagram in figure~\ref{fig:1113b}(b) is related to those of diagram~\ref{fig:1113b}(a) by a Jacobi identity. One may thus use Jacobi identities to eliminate connected colour factors in favour of alternative ones. In the web-mixing language, this corresponds to the fact that the superposition of eigenvectors of unit eigenvalue produces an eigenvector with unit eigenvalue. Indeed, the Jacobi identity between the three diagrams of figure~\ref{fig:1113b} amounts to the following relation between left eigenvectors of $R_{(1,1,1,3)}$:
\begin{equation}
 \left(\begin{array}{r}-1\\1\\0\\-1\\0\\1\end{array}\right)
-\left(\begin{array}{r}-1\\0\\1\\-1\\1\\0\end{array}\right)
=\left(\begin{array}{r}0\\1\\-1\\0\\-1\\1\end{array}\right).
\label{eigeneq}
\end{equation}

\subsubsection{$W_{(1,1,1,2)}^{3g}$} 
\label{sec:w1112}

\begin{figure}[htb]
\begin{center}
\begin{minipage}[b]{0.3\linewidth}
\scalebox{1}{\includegraphics[angle=90]{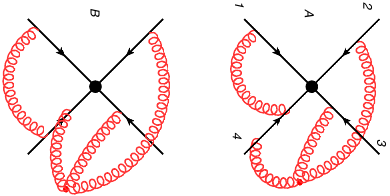}}
\end{minipage}
\begin{minipage}[b]{0.14\linewidth}
\begin{tabular}{c}
\scalebox{.5}{\includegraphics{large_arrow.pdf}}
\\
\\
\\
\\
\\
\\
\\
\\
\\
\\
\\
\\
\,
\end{tabular}
\end{minipage}
\begin{minipage}[b]{0.1\linewidth}
\scalebox{1}{\includegraphics{four_eikonals_1122_connected.pdf}}
\\
\\
\\
\end{minipage}
\vspace*{-80pt}
\caption{The (1,1,1,2) web, with corresponding connected colour factor.}
\label{fig:1112_3g}
\end{center}
\end{figure}

This is characterised in web-mixing matrix language as follows:
\begin{align}
\begin{split}
&r_{(1,1,1,2)}=1\\
&R_{(1,1,1,2)}=\frac16
 \left[ \begin {array}{cc} 3&-3\\ \noalign{\medskip}-3&3\end {array}
 \right] 
\,,\qquad
\left[ \begin {array}{c} A\\
\noalign{\medskip}B
\end{array}\right]
=
 \left[ \begin {array}{c} [[1],[2],[2],[2,1]]\\ \noalign{\medskip}[[1]
,[2],[2],[1,2]]\end {array} \right] 
\end{split}
\end{align}
\begin{align}
\begin{split}
Y_{(1,1,1,2)}= \left[ \begin {array}{cc} -1&1\\ \noalign{\medskip}1&1
\end {array} \right] \,,\qquad
Y_{(1,1,1,2)}^{-1}=\left[ \begin {array}{cc} -1/2&1/2
\\ \noalign{\medskip}1/2&1/2\end {array} \right] .
\end{split}
\end{align}
The rank of this web-mixing matrix is 1 and the single connected colour factor is
\begin{align}
\begin{split}
(YC)_1&=-\ii T_1^aT_2^cT_3^d\Big(-T_4^{ba}+T_4^{ab}\Big) f^{bcd}
=f^{abe}f^{bcd} T_1^a T_2^c T_3^d T_4^e.
\end{split}
\end{align}
This is depicted on the right-hand side of figure~\ref{fig:1112_3g}.  The corresponding
kinematic factor is
\begin{align}
\begin{split}
f_1&=\frac12\Big(-{\cal F}(A)+{\cal F}(B)\Big)\,.
\end{split}
\end{align}

\subsection{Webs connecting three lines}
\label{sec:webs-connecting-four-app}
We now turn to webs that connect three Wilson lines.  These results are
summarised in section~\ref{sec:3lines}.

\subsubsection{$W_{(1,2,3)}$} 
\label{sec:w123}
This web was considered as an example of the use of effective vertices in
section~\ref{sec:effect-vert-pract}.
\begin{figure}[hbt]
\begin{center}
\begin{minipage}[b]{0.55\linewidth}
\begin{tabular}{c}
\scalebox{1}{\includegraphics{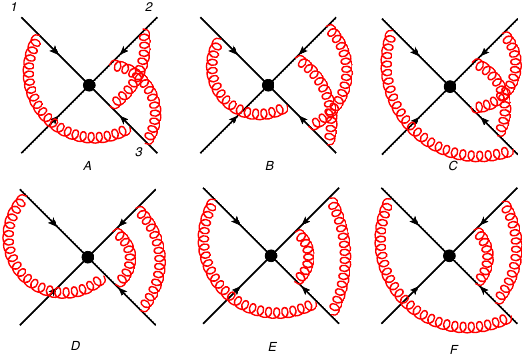}}
\\
\\
\\
\\
\\
\\
\\
\\
\\
\\
\\
\\
\\
\\
\\
\end{tabular}
\end{minipage}
\begin{minipage}[b]{0.15\linewidth}
\begin{tabular}{c}
\hspace*{30pt}\scalebox{.35}{\includegraphics{large_arrow.pdf}}
\\
\\
\\
\\
\\
\\
\\
\\
\\
\\
\\
\\
\\
\\
\,
\end{tabular}
\end{minipage}
\begin{minipage}[b]{0.1\linewidth}
\scalebox{.8}{\includegraphics{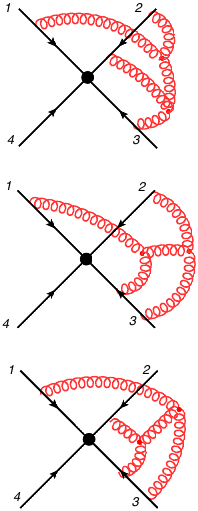}}
\end{minipage}
\vspace*{-180pt}
\caption{The diagrams in the 1-2-3 web with the resulting connected colour
  factors which appear in the exponent.}
\label{fig:123}
\end{center}
\end{figure}
The web-mixing matrix analysis gives
\begin{align}
\label{w123}
\begin{split}
&r_{(1,2,3)}=3\\
&R_{(1,2,3)}=\frac16
 \left[ \begin {array}{cccccc} 6&-3&-3&1&-2&1\\ \noalign{\medskip}0&3&
-3&-2&-2&4\\ \noalign{\medskip}0&-3&3&4&-2&-2\\ \noalign{\medskip}0&0&0
&1&-2&1\\ \noalign{\medskip}0&0&0&-2&4&-2\\ \noalign{\medskip}0&0&0&1&
-2&1\end {array} \right] 
\,,\qquad
\left[ \begin {array}{c} A\\
\noalign{\medskip}B\\
\noalign{\medskip}C\\
\noalign{\medskip}D\\
\noalign{\medskip}E\\
\noalign{\medskip}F
\end{array}\right]
=
 \left[ \begin {array}{c} [[1],[2,3],[3,1,2]]\\ \noalign{\medskip}[[1]
,[2,3],[3,2,1]]\\ \noalign{\medskip}[[1],[2,3],[1,3,2]]
\\ \noalign{\medskip}[[1],[2,3],[2,3,1]]\\ \noalign{\medskip}[[1],[2,3
],[2,1,3]]\\ \noalign{\medskip}[[1],[2,3],[1,2,3]]\end {array}
 \right] 
\end{split}
\end{align}
\begin{align}
\label{eq:yw123}
\begin{split} 
&Y_{(1,2,3)}= \left[ \begin {array}{cccccc} -2&2&0&-1&0&1
\\ \noalign{\medskip}-1&1&0&-1&1&0\\ \noalign{\medskip}-2&1&1&0&0&0
\\ \noalign{\medskip}0&1&1&0&1&0\\ \noalign{\medskip}0&-1/2&-1/2&1&0&0
\\ \noalign{\medskip}0&-1/2&-1/2&0&0&1\end {array} \right] 
\end{split}
\end{align}
\begin{align}
\begin{split}
&Y_{(1,2,3)}^{-1}= \left[ \begin {array}{cccccc} 1/6&-1/3&-1/2&1/3&-1/6&-1/6
\\ \noalign{\medskip}2/3&-1/3&-1/2&1/3&1/3&-2/3\\ \noalign{\medskip}-1
/3&-1/3&1/2&1/3&-2/3&1/3\\ \noalign{\medskip}1/6&-1/3&0&1/3&5/6&-1/6
\\ \noalign{\medskip}-1/3&2/3&0&1/3&1/3&1/3\\ \noalign{\medskip}1/6&-1
/3&0&1/3&-1/6&5/6\end {array} \right] \,.
\end{split}
\end{align}
For this web, we will manipulate the particular choice of basis for the
combinations of colour factors.  For brevity, we will write the $i$-th row of
$Y_{(1,2,3)}$ as $y_i$ such that:
\[
y_1=(-2,2,0,-1,0,1),\qquad y_2=(-1,1,0,-1,1,0),\qquad y_3=(-2,1,1,0,0,0).
\]
We observe that these have entries outside the canonical set ${+1,-1,0}$. However, using a simple rotation in this subspace we obtain three independent eigenvectors with ${+1,-1,0}$ as follows:
\[
u_1=y_1-y_2=(-1,1,0,0,-1,1),\qquad u_2=y_2=(-1,1,0,-1,1,0),\qquad u_3=y_1-y_2-y_3=(1,0,-1,0,-1,1)\,.
\]
We shall now see that these ECFs have a simple interpretation as connected diagrams. Using a matrix notation $U_{ij}=u_i^{(j)}$ and $\left(UC\right)_i=c_i$
\begin{align}
\begin{split}
c_1=(UC)_1&=T_1^aT_2^{bc}\,\Big(-T_3^{cab}+T_3^{cba}-T_3^{bac}+T_3^{abc}\Big)=f^{abd}f^{cde}T_1^aT_2^{bc}T_3^{e}
\\
c_2=(UC)_2&=T_1^aT_2^{bc}\,\Big(-T_3^{cab}+T_3^{cba}-T_3^{bca}+T_3^{bac}\Big)=f^{bce}f^{abd}T_1^aT_2^eT_3^{cd}
\\
c_3=(UC)_3&=T_1^aT_2^{bc}\,\Big(T_3^{cab}-T_3^{acb}-T_3^{bac}+T_3^{abc}\Big)=f^{acd}f^{bce}T_1^aT_2^eT_3^{db}\,.
\end{split}
\end{align}
These are presented  graphically on the right hand side of figure  \ref{fig:123}.

The corresponding combinations of kinematic factors are:
\[
W_{(1,2,3)}=\sum_{i=1}^3 \left({\cal F}Y^{-1}\right)_i \left(YC\right)_i=f_1(c_1+c_2)+f_2c_2+f_3(c_1-c_3)= c_1(f_1+f_3)+c_2(f_1+f_2)-c_3f_3\,.
\]
The kinematic factors are:
\begin{align}
\begin{split}
f_1&=\frac16\Big({\cal F}(A)+4{\cal F}(B)-2{\cal F}(C)+{\cal F}(D)-2{\cal F}(E)+{\cal F}(F)\Big)
\\
f_2&=\frac16\Big(-2{\cal F}(A)-2{\cal F}(B)-2{\cal F}(C)-2{\cal F}(D)+4{\cal F}(E)-2{\cal F}(F)\Big)
\\
f_3&=\frac16\Big(-3{\cal F}(A)-3{\cal F}(B)+3{\cal F}(C)\Big)
\end{split}
\end{align}
so 
\begin{align}
\begin{split}
f_1+f_3&=\frac16\Big(-2{\cal F}(A)+{\cal F}(B)+{\cal F}(C)+{\cal F}(D)-2{\cal F}(E)+{\cal F}(F)\Big)
\\
f_1+f_2&=\frac16\Big(-{\cal F}(A)+2{\cal F}(B)-4{\cal F}(C)-{\cal F}(D)+2{\cal F}(E)-{\cal F}(F)\Big)\,.
\end{split}
\end{align}

\subsubsection{$W_{(1,2,2)}^{3g}$} 
\label{sec:w1223g}
\begin{figure}[htb]
\begin{center}
\begin{minipage}[b]{0.45\linewidth}
\scalebox{.6}{\includegraphics[angle=0]{four_eikonals_122.pdf}}
\end{minipage}
\begin{minipage}[b]{0.15\linewidth}
\begin{tabular}{c}
\scalebox{.4}{\includegraphics{large_arrow.pdf}}
\\
\\
\\
\\
\\
\,
\end{tabular}
\end{minipage}
\begin{minipage}[b]{0.1\linewidth}
\scalebox{0.60}{\includegraphics{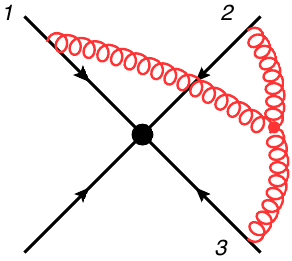}}
\end{minipage}
\vspace*{-40pt}

\caption{A (1,2,2) web with three-gluon vertex, with corresponding connected colour factor.}
\label{fig:122}
\end{center}
\end{figure}
This web is classified by
\begin{align}
\begin{split}
&r_{(1,2,2)}=1\\
&R_{(1,2,2)}=\frac16
 \left[ \begin {array}{cc} 3&-3\\ \noalign{\medskip}-3&3\end {array}
 \right] 
\,,\qquad
\left[ \begin {array}{c} A\\
\noalign{\medskip}B
\end{array}\right]
=
 \left[ \begin {array}{c} [[1],[2,1],[2,2]]\\ \noalign{\medskip}[[1],[
1,2],[2,2]]\end {array} \right] 
\end{split}
\end{align}
\begin{align}
\begin{split}
&
Y_{(1,2,2)}= \left[ \begin {array}{cc} -1&1\\ \noalign{\medskip}1&1
\end {array} \right] \,,\qquad
Y_{(1,2,2)}^{-1}= \left[ \begin {array}{cc} -1/2&1/2
\\ \noalign{\medskip}1/2&1/2\end {array} \right] \,.
\end{split}
\end{align}

\begin{figure}
\begin{center}
\scalebox{0.8}{\includegraphics{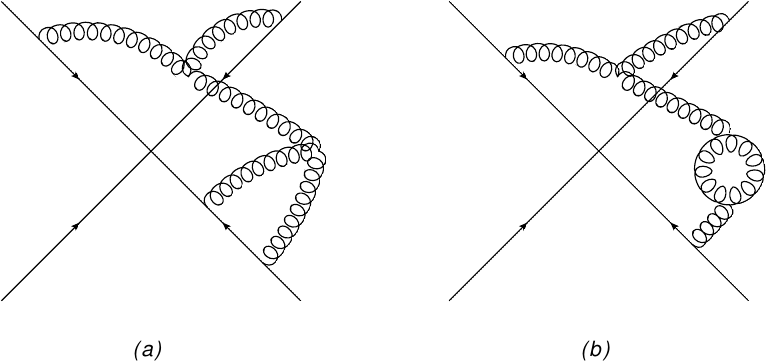}}
\caption{(a) The connected colour factor associated with the (1,2,2) web of
  figure~\ref{fig:122}; (b) an equivalent colour factor.}
\label{fig:colfac122}
\end{center}
\end{figure}

The Feynman rule is $\ii f^{abc}$ for three gluons $a$, $b$ and $c$ going
clockwise and so we find 
\begin{align}
\label{ecf_122}
\begin{split}
(YC)_1&= -\ii f^{abc}T_1^d \Big(-T_2^{bd}+T_2^{db}\Big)T_3^{ac}
= f^{abc}T_1^d  f^{dbe} T_2^e T_3^{ac}\\
&=\frac12 f^{abc}T_1^d  f^{dbe} T_2^e \Big(T_3^{[a,c]}+T_3^{\{a,c\}}\Big)
=\frac{\ii}{2} f^{abc} f^{acg}  f^{dbe}T_1^d T_2^e T_3^g 
\\&=-\ii \frac{N_c}{2}f^{dbe}T_1^d T_2^e T_3^b \,.
\end{split}
\end{align}
The kinematic factor is:
\begin{align}
\label{f1_122}
\begin{split}
f_1&=\frac12\Big(-{\cal F}(A)+{\cal F}(B)\Big).
\end{split}
\end{align}
Note that we have chosen to write the colour factor as being that of a $Y$ graph connecting the three Wilson lines, and an overall normalisation which contains the quadratic Casimir invariant $C_A=N_c$. We could also have chosen to identify the colour factor as that of the graph shown in figure~\ref{fig:colfac122}(a), in which two gluons are emitted from one of the lines. This is equivalent to the colour factor of the graph shown in figure~\ref{fig:colfac122}(b), where one may use antisymmetry of the structure constants to replace the two gluon emissions on the third line with a self-energy bubble. This is the origin of the quadratic Casimir which appears in eq.~(\ref{ecf_122}), and tells us that the connected colour factor graph which appears here corresponds to a one-loop soft gluon scattering topology spanning the three Wilson lines, in contrast to the purely tree-level topologies that appeared when four Wilson lines were connected. This example also tells us that whether or not we consider the connected colour graphs to be tree-level or one-loop is partly a matter of choice, as can be seen by comparing figs.~\ref{fig:colfac122}(a) and (b). \\

\subsubsection{$W_{(1,2,3)}^{\rm SE}$} 
\label{sec:w123se}
\begin{figure}[hbt]
\begin{center}
\begin{minipage}[b]{0.55\linewidth}
\scalebox{.8}{\includegraphics[angle=0]{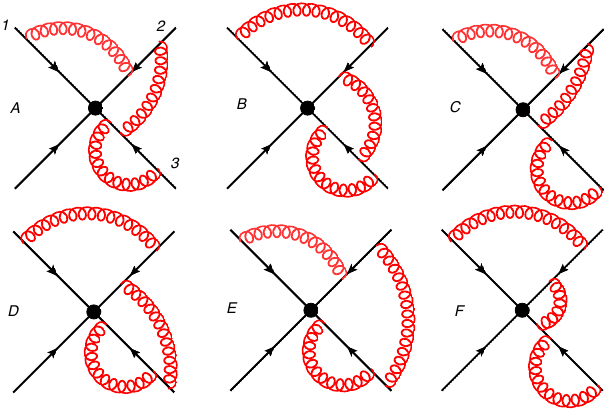}}
\end{minipage}
\begin{minipage}[b]{0.15\linewidth}
\begin{tabular}{c}
\scalebox{.4}{\includegraphics{large_arrow.pdf}}
\\
\\
\\
\\
\\
\\
\\
\\
\\
\,
\end{tabular}
\end{minipage}
\begin{minipage}[b]{0.1\linewidth}
\scalebox{0.67}{\includegraphics{four_eikonals_122_connected.pdf}}
\\
\\
\end{minipage}
\vspace*{-60pt}

\caption{A 1-2-3 web which contains a gluon emitted and absorbed by the same
  Wilson line.}
\label{fig:123_SE}
\end{center}
\end{figure}

\begin{align}
\begin{split}
&r_{(1,2,3)}=2\\
&R_{(1,2,3)}=\frac16
 \left[ \begin {array}{cccccc} 3&-3&-1&2&-2&1\\ \noalign{\medskip}-3&3
&2&-1&1&-2\\ \noalign{\medskip}0&0&2&2&-2&-2\\ \noalign{\medskip}0&0&2
&2&-2&-2\\ \noalign{\medskip}0&0&-1&-1&1&1\\ \noalign{\medskip}0&0&-1&
-1&1&1\end {array} \right] 
\,,\qquad
\left[ \begin {array}{c} A\\
\noalign{\medskip}B\\
\noalign{\medskip}C\\
\noalign{\medskip}D\\
\noalign{\medskip}E\\
\noalign{\medskip}F
\end{array}\right]
=
 \left[ \begin {array}{c} [[1],[2,1],[3,2,3]]\\ \noalign{\medskip}[[1]
,[1,2],[3,2,3]]\\ \noalign{\medskip}[[1],[2,1],[3,3,2]]
\\ \noalign{\medskip}[[1],[1,2],[2,3,3]]\\ \noalign{\medskip}[[1],[2,1
],[2,3,3]]\\ \noalign{\medskip}[[1],[1,2],[3,3,2]]\end {array}
 \right] 
\end{split}
\end{align}
\begin{align}
\begin{split}
&
Y_{(1,2,3)}= \left[ \begin {array}{cccccc} 1&-1&-1&0&0&1
\\ \noalign{\medskip}-1&1&0&-1&1&0\\ \noalign{\medskip}1&1&0&0&0&1
\\ \noalign{\medskip}1&1&0&0&1&0\\ \noalign{\medskip}-2&-2&1&0&0&0
\\ \noalign{\medskip}-2&-2&0&1&0&0\end {array} \right] 
\end{split}
\end{align}
\begin{align}
\begin{split}
&
Y_{(1,2,3)}^{-1}=
\left[ \begin {array}{cccccc} 1/6&-1/3&-1/6&1/3&1/6&-1/3
\\ \noalign{\medskip}-1/3&1/6&1/3&-1/6&-1/3&1/6\\ \noalign{\medskip}-1
/3&-1/3&1/3&1/3&2/3&-1/3\\ \noalign{\medskip}-1/3&-1/3&1/3&1/3&-1/3&2/
3\\ \noalign{\medskip}1/6&1/6&-1/6&5/6&1/6&1/6\\ \noalign{\medskip}1/6
&1/6&5/6&-1/6&1/6&1/6\end {array} \right] .
\end{split}
\end{align}

We therefore find here:
\begin{align}
\begin{split}
(YC)_1&=T_1^a \Big(T_2^{ba}T_3^{cbc}-T_2^{ab}T_3^{cbc}-T_2^{ba}T_3^{ccb}+T_2^{ab}T_3^{ccb}\Big)=\ii \frac{N_c}{2} f^{abd}T_1^aT_2^dT_3^b\\
(YC)_2&=T_1^a \Big(-T_2^{ba}T_3^{cbc}+T_2^{ab}T_3^{cbc}-T_2^{ab}T_3^{bcc}+T_2^{ba}T_3^{bcc}\Big)
=-\ii \frac{N_c}{2} f^{abd}T_1^aT_2^dT_3^b = -(YC)_1\,.
\end{split}
\end{align}
This is a rare example where the rank of the matrix actually overcounts the
number of independent colour factors.  We only encounter this in webs where a
gluon is emitted and absorbed by the same line.  The kinematic factors are:
\begin{align}
\begin{split}
f_1&=\frac16\Big({\cal F}(A)-2{\cal F}(B)-2{\cal F}(C)-2{\cal F}(D)+{\cal F}(E)+{\cal F}(F)\Big)
\\
f_2&=\frac16\Big(-2{\cal F}(A)+{\cal F}(B)-2{\cal F}(C)-2{\cal F}(D)+{\cal F}(E)+{\cal F}(F)\Big)\,,
\end{split}
\end{align}
so the total contribution of this web is
\begin{align}
W^{\rm SE}_{(1,2,3)}=(YC)_1f_1+(YC)_2f_2=
(YC)_1\Big(f_1-f_2\Big)
\end{align}
where 
\[
f_1-f_2=\frac12\Big({\cal F}(A)-{\cal F}(B)\Big)\,.
\]

\subsubsection{$W_{(1,1,3)}^{3g}$} 
\label{sec:w1133g}
\begin{figure}[htb]
\begin{minipage}[b]{0.5\linewidth}
\scalebox{.7}{\includegraphics[angle=0]{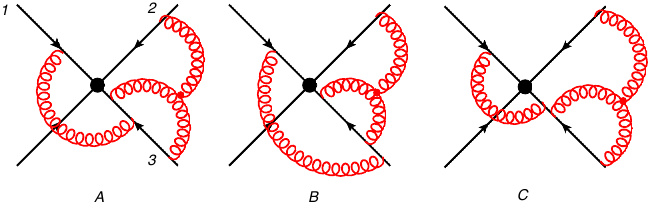}}
\end{minipage}
\begin{minipage}[b]{0.12\linewidth}
\begin{tabular}{c}
\scalebox{.4}{\includegraphics{large_arrow.pdf}}
\\
\\
\\
\\
\\
\,
\end{tabular}
\end{minipage}
\begin{minipage}[b]{0.1\linewidth}
\scalebox{0.7}{\includegraphics{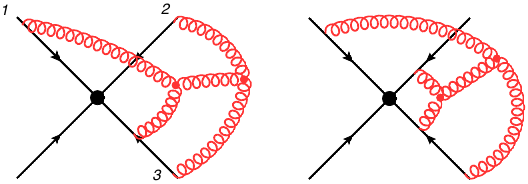}}
\end{minipage}
\vspace*{-30pt}

\caption{A (1,1,3) web with a three-gluon vertex where two of its gluons attach
  to the same line. The resulting connected colour factors are shown on the right.}
\label{fig:113_3g}
\end{figure}

\begin{align}
\begin{split}
&r_{(1,1,3)}=2\\
&R_{(1,1,3)}=\frac16
 \left[ \begin {array}{ccc} 6&-3&-3\\ \noalign{\medskip}0&3&-3
\\ \noalign{\medskip}0&-3&3\end {array} \right] 
\,,\qquad
\left[ \begin {array}{c} A\\
\noalign{\medskip}B\\
\noalign{\medskip}C
\end{array}\right]
=
 \left[ \begin {array}{c} [[1],[2],[2,1,2]]\\ \noalign{\medskip}[[1],[
2],[1,2,2]]\\ \noalign{\medskip}[[1],[2],[2,2,1]]\end {array} \right] 
\end{split}
\end{align}
\begin{align}
\begin{split}
&
Y_{(1,1,3)}= \left[ \begin {array}{ccc} -1&0&1\\ \noalign{\medskip}-1&
1&0\\ \noalign{\medskip}0&1&1\end {array} \right] \,,\qquad
Y_{(1,1,3)}^{-1}= \left[ \begin {array}{ccc} -1/2&-1/2&1/2
\\ \noalign{\medskip}-1/2&1/2&1/2\\ \noalign{\medskip}1/2&-1/2&1/2
\end {array} \right] \,.
\end{split}
\end{align}

The two connected colour factors are found to be
\begin{align}
\begin{split}
(YC)_1&=-\ii f^{bcd} T_1^a T_2^b\,\Big(-T_3^{dac}+T_3^{dca}\Big)
= f^{bcd}f^{cae}T_1^aT_2^bT_3^{de}
\\
(YC)_2&=-\ii f^{bcd} T_1^a T_2^b\,\Big(-T_3^{dac}+T_3^{adc}\Big)
= f^{bcd}f^{ade}T_1^aT_2^bT_3^{ec}
\end{split}
\end{align}
with kinematic coefficients
\begin{align}
\begin{split}
f_1&=\frac12\Big(-{\cal F}(A)-{\cal F}(B)+{\cal F}(C)\Big)
\\
f_2&=\frac12\Big(-{\cal F}(A)+{\cal F}(B)-{\cal F}(C)\Big).
\end{split}
\end{align}

\subsubsection{$W_{(1,1,4)}^{\rm SE}$} 
\label{sec:w114se}

This is the only web we encounter with four gluon attachments on a single line.
The twelve diagrams are shown with their labels in fig.~\ref{fig:114_SE}.  We find:
\begin{figure}[hbt]
\begin{center}
\begin{minipage}[b]{0.6\linewidth}
\scalebox{.7}{\includegraphics[angle=0]{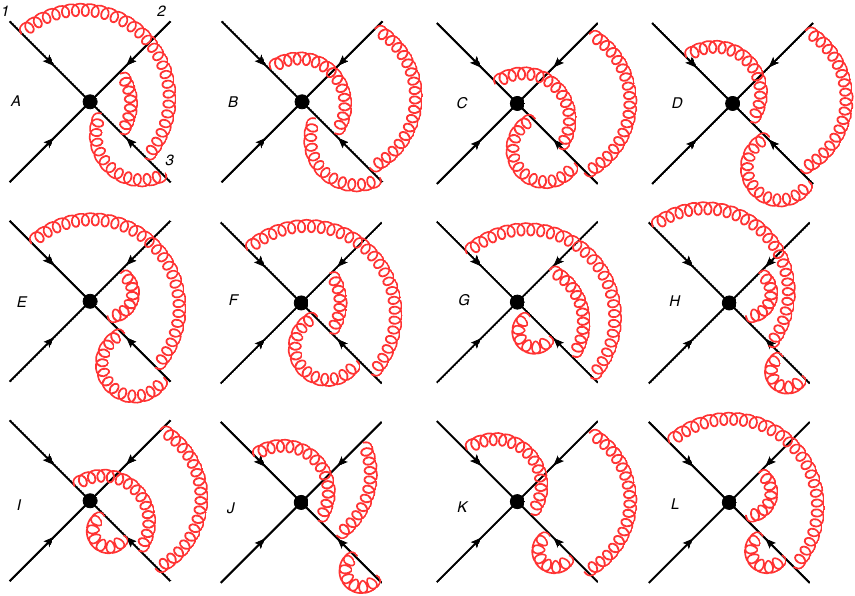}}
\end{minipage}
\begin{minipage}[b]{0.12\linewidth}
\begin{tabular}{c}
\scalebox{.4}{\includegraphics{large_arrow.pdf}}
\\
\\
\\
\\
\\
\\
\\
\\
\\
\\
\\
\\
\,
\end{tabular}
\end{minipage}
\begin{minipage}[b]{0.1\linewidth}
\scalebox{0.65}{\includegraphics{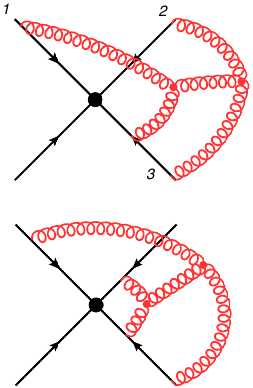}}
\\
\\
\end{minipage}
\vspace*{-80pt}
\caption{The diagrams for the 1-1-4 web and the resulting connected colour
  factor.  Note that these are exactly the same two as in figure \ref{fig:113_3g}.}
\label{fig:114_SE}
\end{center}
\end{figure}

\begin{align}
\begin{split}
&r_{(1,1,4)}=6\\
&R_{(1,1,4)}=\frac16
 \left[ \begin {array}{cccccccccccc} 6&0&-3&-3&-3&-3&-1&-1&2&2&2&2
\\ \noalign{\medskip}0&6&-3&-3&-3&-3&2&2&-1&-1&2&2
\\ \noalign{\medskip}0&0&3&0&-3&0&-1&2&-1&-1&-1&2\\ \noalign{\medskip}0
&0&0&3&0&-3&2&-1&-1&-1&-1&2\\ \noalign{\medskip}0&0&-3&0&3&0&-1&-1&2&-
1&2&-1\\ \noalign{\medskip}0&0&0&-3&0&3&-1&-1&-1&2&2&-1
\\ \noalign{\medskip}0&0&0&0&0&0&2&-1&-1&2&-1&-1\\ \noalign{\medskip}0
&0&0&0&0&0&-1&2&2&-1&-1&-1\\ \noalign{\medskip}0&0&0&0&0&0&-1&2&2&-1&-
1&-1\\ \noalign{\medskip}0&0&0&0&0&0&2&-1&-1&2&-1&-1
\\ \noalign{\medskip}0&0&0&0&0&0&-1&-1&-1&-1&2&2\\ \noalign{\medskip}0
&0&0&0&0&0&-1&-1&-1&-1&2&2\end {array} \right] 
\end{split}
\end{align}
\begin{align}
\begin{split}
&
\left[ \begin {array}{c} A\\
\noalign{\medskip}B\\
\noalign{\medskip}C\\
\noalign{\medskip}D\\
\noalign{\medskip}E\\
\noalign{\medskip}F\\
\noalign{\medskip}G\\
\noalign{\medskip}H\\
\noalign{\medskip}I\\
\noalign{\medskip}J\\
\noalign{\medskip}K\\
\noalign{\medskip}L
\end{array}\right]
=
 \left[ \begin {array}{c} [[1],[2],[3,1,2,3]]\\ \noalign{\medskip}[[1]
,[2],[3,2,1,3]]\\ \noalign{\medskip}[[1],[2],[2,3,1,3]]
\\ \noalign{\medskip}[[1],[2],[3,2,3,1]]\\ \noalign{\medskip}[[1],[2],
[3,1,3,2]]\\ \noalign{\medskip}[[1],[2],[1,3,2,3]]
\\ \noalign{\medskip}[[1],[2],[1,2,3,3]]\\ \noalign{\medskip}[[1],[2],
[3,3,1,2]]\\ \noalign{\medskip}[[1],[2],[2,1,3,3]]
\\ \noalign{\medskip}[[1],[2],[3,3,2,1]]\\ \noalign{\medskip}[[1],[2],
[2,3,3,1]]\\ \noalign{\medskip}[[1],[2],[1,3,3,2]]\end {array}
 \right] 
\end{split}
\end{align}
\begin{align}
\begin{split}
&
Y_{(1,1,4)}= \left[ \begin {array}{cccccccccccc} 1&0&0&-1&-1&0&0&0&0&1
&0&0\\ \noalign{\medskip}0&1&-1&-1&0&0&0&0&0&0&1&0
\\ \noalign{\medskip}1&0&0&0&-1&-1&0&0&0&0&0&1\\ \noalign{\medskip}0&1
&0&-1&-1&0&0&1&0&0&0&0\\ \noalign{\medskip}1&0&-1&0&0&-1&0&0&1&0&0&0
\\ \noalign{\medskip}0&1&-1&0&0&-1&1&0&0&0&0&0\\ \noalign{\medskip}0&0
&0&1&0&1&0&1&0&0&0&0\\ \noalign{\medskip}0&0&1&0&1&0&1&0&0&0&0&0
\\ \noalign{\medskip}0&0&0&1&0&1&0&0&1&0&0&0\\ \noalign{\medskip}0&0&1
&0&1&0&0&0&0&1&0&0\\ \noalign{\medskip}0&0&-1&-1&-1&-1&0&0&0&0&1&0
\\ \noalign{\medskip}0&0&-1&-1&-1&-1&0&0&0&0&0&1\end {array} \right] 
\end{split}
\end{align}
and finally
\begin{align}
\begin{split}
&
Y_{(1,1,4)}^{-1}=
\left[ \begin {array}{cccccccccccc} 1/3&1/3&1/3&-1/6&1/3&-1/6&1/6&1/6
&-1/3&-1/3&-1/3&-1/3\\ \noalign{\medskip}-1/6&1/3&1/3&1/3&-1/6&1/3&-1/
3&-1/3&1/6&1/6&-1/3&-1/3\\ \noalign{\medskip}-1/6&-1/6&1/3&1/3&-1/6&-1
/6&-1/3&1/6&1/6&1/6&1/6&-1/3\\ \noalign{\medskip}-1/6&-1/6&1/3&-1/6&-1
/6&1/3&1/6&-1/3&1/6&1/6&1/6&-1/3\\ \noalign{\medskip}-1/6&1/3&-1/6&-1/
6&1/3&-1/6&1/6&1/6&-1/3&1/6&-1/3&1/6\\ \noalign{\medskip}1/3&1/3&-1/6&
-1/6&-1/6&-1/6&1/6&1/6&1/6&-1/3&-1/3&1/6\\ \noalign{\medskip}1/3&-1/6&
-1/6&-1/6&-1/6&1/3&1/6&2/3&1/6&-1/3&1/6&1/6\\ \noalign{\medskip}-1/6&-
1/6&-1/6&1/3&1/3&-1/6&2/3&1/6&-1/3&1/6&1/6&1/6\\ \noalign{\medskip}-1/
6&-1/6&-1/6&1/3&1/3&-1/6&-1/3&1/6&2/3&1/6&1/6&1/6\\ \noalign{\medskip}
1/3&-1/6&-1/6&-1/6&-1/6&1/3&1/6&-1/3&1/6&2/3&1/6&1/6
\\ \noalign{\medskip}-1/6&1/3&1/3&-1/6&-1/6&-1/6&1/6&1/6&1/6&1/6&2/3&-
1/3\\ \noalign{\medskip}-1/6&1/3&1/3&-1/6&-1/6&-1/6&1/6&1/6&1/6&1/6&-1
/3&2/3\end {array} \right] \,.
\end{split}
\end{align}

Naming the diagrams as labelled on the left-hand-side of fig.~\ref{fig:114_SE}, we get the following ECFs:
\begin{align}
\begin{split}
(YC)_1&= A-D-E+J=T_1^a T_2^b\,\Big(T_3^{cabc}-T_3^{cbca}-T_3^{cacb}+T_3^{ccba}\Big)
=f^{bdc}f^{ade}T_1^aT_2^bT_3^{ce}
\\
(YC)_2&= B-C-D+K=T_1^a T_2^b\,\Big(T_3^{cbac}-T_3^{bcac}-T_3^{cbca}+T_3^{bcca}\Big)
=(YC)_1
\\
(YC)_3&=A-E-F+L=T_1^a T_2^b\,\Big(T_3^{cabc}-T_3^{cacb}-T_3^{acbc}+T_3^{accb}\Big)
=f^{bcd}f^{ace}T_1^aT_2^bT_3^{ed}
\\
(YC)_4&=B-D-E+H=T_1^a T_2^b\,\Big(T_3^{cbac}-T_3^{cbca}-T_3^{cacb}+T_3^{ccab}\Big)
=(YC)_3
\\
(YC)_5&=A-C-F+I=T_1^a T_2^b\,\Big(T_3^{cabc}-T_3^{bcac}-T_3^{acbc}+T_3^{bacc}\Big)
=f^{adc}f^{bde}T_1^aT_2^bT_3^{ec}=(YC)_1
\\
(YC)_6&=B-C-F+G=T_1^a T_2^b\,\Big(T_3^{cbac}-T_3^{bcac}-T_3^{acbc}+T_3^{abcc}\Big)
= f^{cdb}f^{eda} T_1^a T_2^b\,T_3^{ec}=(YC)_3.
\end{split}
\end{align}
Again we see an example of the rank over-counting the number of independent
colour factors where the diagram involves a gluon absorbed and emitted from the
same Wilson line.  The kinematic factors are (for brevity we omit here the
notation ${\cal F}$):
\begin{align}
\begin{split}
f_1&=\frac16\Big(2A-B-C-D-E+2F+2G-H-I+2J-K-L\Big)
\\
f_2&=\frac16\Big(2A+2B-C-D+2E+2F-G-H-I-J+2K+2L\Big)
\\
f_3&=\frac16\Big(2A+2B+2C+2D-E-F-G-H-I-J+2K+2L\Big)
\\
f_4&=\frac16\Big(-A+2B+2C-D-E-F-G+2H+2I-J-K-L\Big)
\\
f_5&=\frac16\Big(2A-B-C-D+2E-F-G+2H+2I-J-K-L\Big)
\\
f_6&=\frac16\Big(-A+2B-C+2D-E-F+2G-H-I+2J-K-L\Big)\,.
\end{split}
\end{align}
Overall we get for this web
\begin{align}
\begin{split}
W_{(1,1,4)}^{\rm SE} &= (YC)_1 f_1+ (YC)_2 f_2+(YC)_3 f_3+(YC)_4 f_4+(YC)_5 f_5+(YC)_6 f_6
\\
&= (YC)_1 \Big(f_1+  f_2 + f_5\Big)  +(YC)_3 \Big( f_3+ f_4+ f_6\Big)
\end{split}
\end{align}
where we find:
\begin{align}
\begin{split}
f_1+  f_2 + f_5&=\frac12\left(2A-C-D+E+F\right)\\
f_3+f_4+f_6&=\frac12\left(2B+C+D-E-F\right)\,.
\end{split}
\end{align}

\subsubsection{$W_{(2,2,2)}$} 
\label{sec:w222}

\begin{figure}[htb]
\begin{minipage}[b]{0.52\linewidth}
\scalebox{.6}{\includegraphics[angle=0]{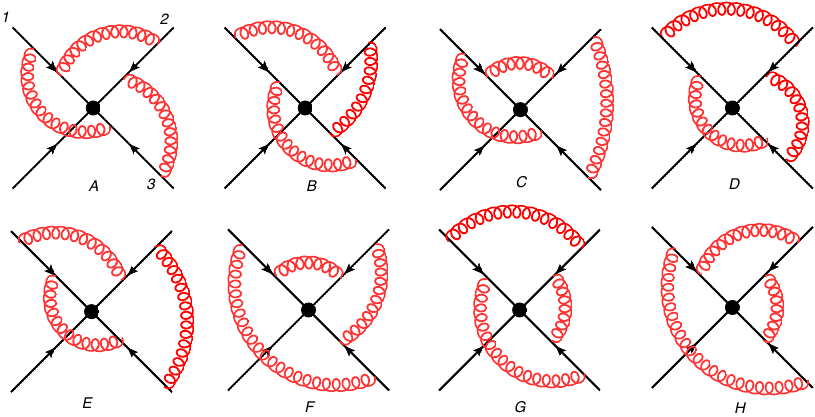}}
\end{minipage}
\begin{minipage}[b]{0.13\linewidth}
\begin{tabular}{c}
\scalebox{.43}{\includegraphics{large_arrow.pdf}}
\\
\\
\\
\\
\\
\\
\\
\\
\,
\end{tabular}
\end{minipage}
\begin{minipage}[b]{0.1\linewidth}
\scalebox{0.66}{\includegraphics{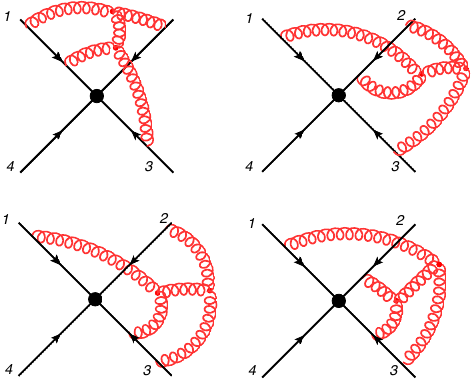}}
\end{minipage}
\vspace*{-60pt}

\caption{A (2,2,2) web, with corresponding connected colour factors.}
\label{fig:222}
\end{figure}

\begin{align}
\begin{split}
&r_{(2,2,2)}=4\\
&R_{(2,2,2)}=\frac16
 \left[ \begin {array}{cccccccc} 6&0&-4&-4&2&2&2&-4
\\ \noalign{\medskip}0&6&2&2&-4&-4&-4&2\\ \noalign{\medskip}0&0&2&-1&-
1&-1&2&-1\\ \noalign{\medskip}0&0&-1&2&-1&2&-1&-1\\ \noalign{\medskip}0
&0&-1&-1&2&-1&-1&2\\ \noalign{\medskip}0&0&-1&2&-1&2&-1&-1
\\ \noalign{\medskip}0&0&2&-1&-1&-1&2&-1\\ \noalign{\medskip}0&0&-1&-1
&2&-1&-1&2\end {array} \right] 
\,,\qquad
\left[ \begin {array}{c} A\\
\noalign{\medskip}B\\
\noalign{\medskip}C\\
\noalign{\medskip}D\\
\noalign{\medskip}E\\
\noalign{\medskip}F\\
\noalign{\medskip}G\\
\noalign{\medskip}H
\end{array}\right]
=
 \left[ \begin {array}{c} [[1,2],[2,3],[3,1]]\\ \noalign{\medskip}[[1,
2],[3,1],[2,3]]\\ \noalign{\medskip}[[1,2],[3,2],[3,1]]
\\ \noalign{\medskip}[[1,2],[1,3],[3,2]]\\ \noalign{\medskip}[[1,2],[3
,1],[3,2]]\\ \noalign{\medskip}[[1,2],[3,2],[1,3]]
\\ \noalign{\medskip}[[1,2],[1,3],[2,3]]\\ \noalign{\medskip}[[1,2],[2
,3],[1,3]]\end {array} \right] 
\end{split}
\end{align}
\begin{align}
\begin{split}
&
Y_{(2,2,2)}= 
\left[ \begin {array}{cccccccc} -1&-1&1&-1&1&-1&1&1
\\ \noalign{\medskip}-1&-1&-1&1&1&1&-1&1\\ \noalign{\medskip}1&1&-1&-1
&1&-1&-1&1\\ \noalign{\medskip}-1&1&1&1&-1&-1&-1&1
\\ \noalign{\medskip}0&0&1&1&0&0&0&1\\ \noalign{\medskip}0&0&-1&0&0&0&
1&0\\ \noalign{\medskip}0&0&0&-1&0&1&0&0\\ \noalign{\medskip}0&0&1&1&1
&0&0&0\end {array} \right] 
\end{split}
\end{align}

\begin{align}
\begin{split}
&
Y_{(2,2,2)}^{-1}=
\left[ \begin {array}{cccccccc} -1/6&-1/6&1/6&-1/2&2/3&-1/3&-1/3&-1/3
\\ \noalign{\medskip}-1/6&-1/6&1/6&1/2&-1/3&2/3&2/3&2/3
\\ \noalign{\medskip}1/12&-1/6&-1/12&0&1/6&-1/3&1/6&1/6
\\ \noalign{\medskip}-1/6&1/12&-1/12&0&1/6&1/6&-1/3&1/6
\\ \noalign{\medskip}1/12&1/12&1/6&0&-1/3&1/6&1/6&2/3
\\ \noalign{\medskip}-1/6&1/12&-1/12&0&1/6&1/6&2/3&1/6
\\ \noalign{\medskip}1/12&-1/6&-1/12&0&1/6&2/3&1/6&1/6
\\ \noalign{\medskip}1/12&1/12&1/6&0&2/3&1/6&1/6&-1/3\end {array}
 \right] 
\end{split}
\end{align}
Denoting the diagrams by $A$ through $H$ as in fig.~\ref{fig:222}, we obtain the four linear combinations of colour factors corresponding to the basis of eq.~(\ref{eq:3legcolour}):
\begin{align}
\begin{split}
(YC)_1&=-A-B+C-D+E-F+G+H 
\\&=\{T_1^{\alpha},T_1^{\beta}\}[T_2^{\beta},T_2^{\gamma}][T_3^{\alpha},T_3^{\gamma}]
\equiv c_1^{ (3)}
\\
(YC)_2&=-A-B-C+D+E+F-G+H 
\\&=[T_1^{\alpha},T_1^{\beta}]\{T_2^{\beta},T_2^{\gamma}\}[T_3^{\alpha},T_3^{\gamma}]
\equiv c_2^{ (3)}
\\
(YC)_3&=A+B-C-D+E-F-G+H 
\\&=  [T_1^{\alpha},T_1^{\beta}][T_2^{\beta},T_2^{\gamma}]\{T_3^{\alpha},T_3^{\gamma}\}
\equiv  c_3^{ (3)}
\\
(YC)_4&=-A+B+C+D-E-F-G+H 
\\&= [T_1^{\alpha},T_1^{\beta}][T_2^{\beta},T_2^{\gamma}][T_3^{\alpha},T_3^{\gamma}]
\equiv c_4^{ (3)}\,.
\end{split}
\end{align}

Thus the total contribution of this web is
\begin{align}
\begin{split}
W_{(2,2,2)}&=(YC)_1f_1+(YC)_2f_2+(YC)_3f_3+(YC)_4f_4\,,
\end{split}
\end{align}
where the kinematic factors are (for brevity we omit here the notation ${\cal F}$, so here $A$ through $F$ correspond to the kinematic factors):
\begin{align}
\begin{split}
f_1&=\frac{1}{12}\Big(-2A-2B+C-2D+E-2F+G+H\Big)
\\
f_2&=\frac{1}{12}\Big(-2A-2B-2C+D+E+F-2G+H\Big)
\\
f_3&=\frac{1}{12}\Big(2A+2B-C-D+2E-F-G+2H\Big)
\\
f_4&=\frac12\Big(-A+B\Big)\,.
\end{split}
\end{align}

\subsubsection{$W_{(1,2,2)}^{3g\prime}$} 
\label{sec:w1223gp}

\begin{figure}[hbt]
\centering
\begin{minipage}[b]{0.52\linewidth}
\scalebox{.6}{\includegraphics[angle=0]{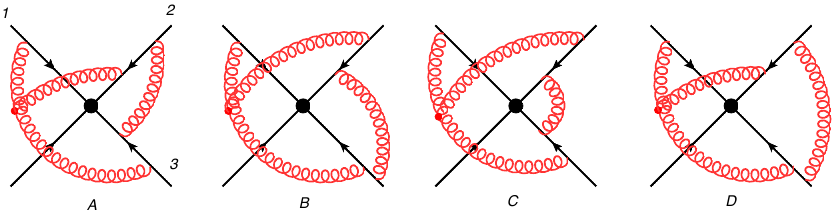}}
\vspace*{1.8cm}
\end{minipage}
\begin{minipage}[b]{0.1\linewidth}
\begin{tabular}{c}
\scalebox{.3}{\includegraphics{large_arrow.pdf}}
\vspace*{5cm}
\,
\end{tabular}
\end{minipage}
\begin{minipage}[b]{0.1\linewidth}
\scalebox{0.6}{\includegraphics{four_eikonals_123_connected.pdf}}
\end{minipage}
\vspace*{-70pt}

\caption{A 1-2-2 web with a three-gluon vertex connecting three different lines
  and its corresponding connected colour factors.}
\label{fig:122_3g}
\end{figure}

\begin{align}
\begin{split}
&r_{(1,2,2)}=3\\
&R_{(1,2,2)}=\frac16
 \left[ \begin {array}{cccc} 6&0&-3&-3\\ \noalign{\medskip}0&6&-3&-3
\\ \noalign{\medskip}0&0&3&-3\\ \noalign{\medskip}0&0&-3&3\end {array}
 \right] 
\,,\qquad
\left[ \begin {array}{c} A\\
\noalign{\medskip}B\\
\noalign{\medskip}C\\
\noalign{\medskip}D
\end{array}\right]
=
 \left[ \begin {array}{c} [[1],[2,1],[1,2]]\\ \noalign{\medskip}[[1],[
1,2],[2,1]]\\ \noalign{\medskip}[[1],[1,2],[1,2]]\\ \noalign{\medskip}
[[1],[2,1],[2,1]]\end {array} \right] 
\end{split}
\end{align}

\begin{align}
\begin{split}
Y_{(1,2,2)}= \left[ \begin {array}{cccc} -1&0&0&1\\ \noalign{\medskip}
-1&0&1&0\\ \noalign{\medskip}-1&1&0&0\\ \noalign{\medskip}0&0&1&1
\end {array} \right] \,,\qquad
&Y_{(1,2,2)}^{-1}= 
\left[ \begin {array}{cccc} -1/2&-1/2&0&1/2
\\ \noalign{\medskip}-1/2&-1/2&1&1/2\\ \noalign{\medskip}-1/2&1/2&0&1/
2\\ \noalign{\medskip}1/2&-1/2&0&1/2\end {array} \right] \,.
\end{split}
\end{align}
We therefore find the colour factors to be
\begin{align}
\begin{split}
(YC)_1&=-\ii T_1^a f^{abc}\Big(-  T_2^{dc}T_3^{bd}+   T_2^{dc}T_3^{db}\Big)
= f^{abc}f^{dbe} T_1^aT_2^{dc}T_3^e\\
(YC)_2&=-\ii T_1^a f^{abc}\Big(-  T_2^{dc}T_3^{bd}+   T_2^{cd}T_3^{bd}\Big)
= f^{abc}f^{cde} T_1^aT_2^{e}T_3^{bd}\\
(YC)_3&=-\ii T_1^a f^{abc}\Big(-  T_2^{dc}T_3^{bd}+   T_2^{cd}T_3^{db}\Big)
= f^{abc} f^{dbe} T_1^aT_2^{dc}T_3^{e} + f^{abc} f^{cde} T_1^aT_2^{e}T_3^{db}\,.
\end{split}
\end{align}
The kinematic factors are:
\begin{align}
\begin{split}
f_1&=\frac12\Big(-{\cal F}(A)-{\cal F}(B)-{\cal F}(C)+{\cal F}(D)\Big)
\\
f_2&=\frac12\Big(-{\cal F}(A)-{\cal F}(B)+{\cal F}(C)-{\cal F}(D)\Big)
\\
f_3&={\cal F}(B)\,.
\end{split}
\end{align}

\bibliography{refs.bib}
\end{document}